\documentclass[aps,prd,preprint,groupedaddress,showpacs]{revtex4}

\usepackage{epsfig}
\usepackage{graphics}
\usepackage{slashed}
\usepackage{color}
\begin{document}

\leftmargin -2cm
\def\choosen{\atopwithdelims..}

\boldmath
\title{$B{\bar B}$ angular correlations at the LHC in parton Reggeization approach merged with higher-order matrix elements}
\unboldmath

\author{\firstname{A.V.}\surname{Karpishkov}} \email{karpishkov@rambler.ru}
\author{\firstname{M.A.}
\surname{Nefedov}} \email{nefedovma@gmail.com}
\author{\firstname{V.A.}\surname{Saleev}} \email{saleev@samsu.ru}

\affiliation{Samara National Research University, Moscow Highway,
34, 443086, Samara, Russia}

\begin{abstract}
We calculate the angular distribution spectra between beauty ($B$)
and anti-beauty ($\bar B$) mesons  in proton-proton collisions in
the leading order approximation of the parton Reggeization approach consistently merged
with the next-to-leading order corrections from the emission of
additional hard gluon. To describe b-quark
hadronization we use  the universal scale-depended parton-to-meson
fragmentation functions extracted from the world $e^+e^-$
annihilation data. We have obtained good agreement between our
predictions and data from the CMS Collaboration at the energy
$\sqrt{S}=7$ TeV for $B \bar B$ angular correlations within
uncertainties and without free parameters. Predictions for analogous correlation observables at $\sqrt{S}=13$ TeV are provided.
\end{abstract}
\pacs{12.38.-t,12.40.Nn,13.85.Ni,14.40.Lb}

\maketitle

\section{Introduction}

%Physics of beauty quarks is very interesting for the search of new exotic bound states of heavy quarks and for precise study parameters of flavour sector of the Standard Model.
Production of $b-$quarks in the high energy $pp-$collisions is the
object of an intensive experimental study at the CERN LHC. In the
present paper we focus on a measurements of $b\bar b$ angular and momentum correlations, since
they provide a test of dynamics of hard interactions, which is
highly sensitive to the higher-order corrections in QCD. There are
two ways of study these $b\bar b$ correlations. The first one is
based on reconstruction of pairs of $b-$jets \cite{bCDF,bATLAS}, in
the second case we get information on dynamics of hard production of
$b\bar b-$pair using data on pair production of $B$-mesons. In turn,
long-lived $B-$mesons are reconstructed via their semileptonic
decays. One advantage of the latter method is the unique capability
to detect $B\bar B-$pairs even at small opening angles, in which
case the decay products of the $B-$hadrons tend to be merged into a
single jet and the standard $b-$jet tagging techniques are not
applicable \cite{bCMS}.

On the theory side, one has to take into account multiple radiation
of both soft/collinear and hard additional partons to describe such
angular correlations over the whole observable range of opening
angles between momenta of $B$-mesons. In the Leading Order (LO) of
Collinear Parton Model (CPM), $b$-quarks are produced back-to-back
in azimuthal angle. Effects of the soft and collinear Initial State
Radiation (ISR) or Final State Radiation (FSR) somewhat smear the
distribution in azimuthal angle difference between transverse
momenta of mesons ($\Delta \phi$) around $\Delta\phi\simeq \pi$.
These effects are systematically taken into account with the Leading
Logarithmic Accuracy in the Parton Showers (PS) of the standad
Monte-Carlo (MC) generators, such as PYTHIA or HERWIG.

Radiation of the additional hard gluons or quarks cause $B$ and
$\bar{B}$ mesons to fly with $\Delta\phi < \pi$, but such radiation
is beyond the formal accuracy of standard PS. Description of such
events essentially depends on the way, how transverse momentum and
the ``small'' light-cone component of momentum of the emitted parton
are dealt with inside a PS algorithm, so called {\it recoiling
scheme}~\cite{PYTHIA_RS}. Usually the accuracy of description of
such kinematic configurations is improved via different methods of
{\it matching} of the full NLO corrections in CPM with the
parton-shower, such as MC@NLO~\cite{MCNLO} or POWHEG~\cite{POWHEG}
or via {\it merging} of the kinematically and dynamically accurate
description of a few additional hard emissions, provided by the
exact tree-level matrix elements, with the soft/collinear emissions
from the PS~\cite{Merging}.

The presence of additional free parameters in the matching/merging
methods, as well as the multitude of possible recoiling schemes,
clearly calls for the improved understanding of the high-$p_T$
regime of the PS from the point of view of Quantum Field Theory.
Apart from the soft and collinear limits, the only known limit of
scattering amplitudes in QCD which structure is sufficiently simple
for the theoretical analysis is the limit of Multi-Regge kinematics
(MRK), when emitted partons are highly separated in rapidity from
each-other. This makes the MRK-limit to be a natural starting point
for the construction of improved approximations. In the present
paper, we construct the factorization formula and the framework of
LO calculations in the Parton Reggeization Approach (PRA), which
unifies the PS-like description of the soft and collinear emissions
with the MRK limit for hard emissions. Then we switch to the
description of the angular correlations in the production of
$B\bar{B}$-pairs accompanied by the hard jet, which sets the scale
of the process. The present study is motivated by experimental data
of the Ref.~\cite{bCMS}, since neither MC-calculations in the
experimantal paper, nor the calculations in the LO of
$k_T$-factorization approach in the Ref.~\cite{JZL_kT-fact} could
accurately describe the shape of angular distributions. We construct
the consistent prescription, which {\it merges} the LO PRA
calculation for this process with tree-level NLO matrix element. The
latter improves description of those events, in which not the
$b$-jet, but the hard gluon jet is the leading one, while avoiding
possible double-counting and divergence problems. In such a way we
achieve a good description of the shape of all $B\bar{B}$
correlation spectra without additional free parameters.

The paper has following structure. We describe the basics of PRA and
it's relationschips with other approaches in the Sec.~\ref{sec:LO}.
In the Sec.~\ref{sec:merging} we present our merging prescription
and the analytic and numerical tools, which we use. Then we
concentrate on the numerical results, comparison with experimental
data of the Ref.~\cite{bCMS} and predictions for possible future
measurements in the Sec.~\ref{sec:results}. Finally, we summarize
our conclusions in the Sec.~\ref{sec:conclusions}.

\section{LO PRA framework}
\label{sec:LO}

To derive the factorization formula of the PRA in LO
approximation, let us consider production of the partonic
final state of interest ${\cal Y}$ in the following auxilliary hard
subprocess:
  \begin{equation}
  g(p_1)+g(p_2)\to g(k_1) + {\cal Y}(P_{\cal A}) + g(k_2), \label{eqI:ax_proc}
  \end{equation}
where the four-momenta of particles are denoted in parthenses, and
$p_1^2=p_2^2=k_1^2=k_2^2=0$. The final state ${\cal Y}$ sets the
hard scale $\mu^2$ of the whole process via it's invariant mass
$M_{\cal A}^2=P_{\cal A}^2$, or transverse momentum $P_{T{\cal A}}$,
otherwise it can be arbitrary combination of QCD partons. In a
frame, where ${\bf p}_{1}=-{\bf p}_{2}$ directed along the
Z-axis it is natural to work with the Sudakov(light-cone) components
of any four-momentum $k$:
  \[
  k^\mu=\frac{1}{2}\left( k^+ n_-^\mu + k^- n_+^\mu \right) + k_T^\mu,
  \]
where $n_{\pm}^\mu = \left(n^\pm\right)^\mu = \left(1,0,0,\mp 1
\right)^\mu$, $n_{\pm}^2=0$, $n_+n^-=2$,
$k^\pm=k_{\pm}=(n_{\pm}k)=k^0\pm k^3$, $n_{\pm}k_T=0$, so that
$p_1^-=p_2^+=0$ and $s=(p_1+p_2)^2=p_1^+p_2^- >0$. The dot-product
of two four-vectors $k$ and $q$ in this notation is equal to:
  \[
  (kq)=\frac{1}{2}\left( k^+ q_- + k^- q_+ \right) - {\bf k}_T {\bf q}_T.
  \]
For the discussion of different kinematic limits of the process
(\ref{eqI:ax_proc}) it is convinient to introduce the
``$t$-channel'' momentum transfers $q_{1,2}=p_{1,2}-k_{1,2}$, which
implies that ${\bf q}_{T1,2}=-{\bf k}_{T1,2}$, $q_{1}^-=-k_1^-$ and
$q_2^+=-k_2^+$. Let us define $t_{1,2}={\bf q}_{T1,2}^2$, and the
corresponding fractions of the ``large'' light-cone components of
momenta:
  \[
  z_1=\frac{q_1^+}{p_1^+},\ \ z_2=\frac{q_2^-}{p_2^-},
  \]
for the further use. Variables $z_{1,2}$ satisfy the conditions
$0\leq z_{1,2}\leq 1$ because  $k_{1,2}^\pm\geq 0$ and
$q_1^+=P_{\cal A}^++k_2^+\geq 0$, $q_2^-=P_{\cal A}^-+k_1^-\geq 0$
since all final-state particles are on-shell.

In the {\it collinear limit} (CL), when ${\bf
k}_{T1,2}^2\ll \mu^2$, while $0\leq z_{1,2}\leq 1$, the asymptotic
for the square of tree-level matrix element for the subprocess
(\ref{eqI:ax_proc}) is very well known:
 \begin{equation}
\overline{|{\cal M}|^2}_{\rm CL} \simeq \frac{4g_s^4}{{\bf
k}_{T1}^2 {\bf k}_{T2}^2} P_{gg}(z_1) P_{gg}(z_2)
\frac{\overline{|{\cal A}_{CPM}|^2}}{z_1 z_2}, \label{eqI:CL_M2}
 \end{equation}
where the bar denotes averaging (summation) over the spin and color
quantum numbers of the initial(final)-state partons, $g_s=\sqrt{4\pi
\alpha_s}$ is the coupling constant of QCD, $P_{gg}(z)=2C_A\left(
(1-z)/z + z/(1-z)+ z(1-z) \right)$ is the LO gluon-gluon DGLAP
splitting function and ${\cal A}_{CPM}$ is the amplitude of the
subprocess $g(z_1p_1)+g(z_2p_2)\to {\cal Y}(P_{\cal A})$ with
on-shell initial-state gluons. The error of approximation
(\ref{eqI:CL_M2}) is suppressed as $O({\bf k}_{T1,2}^2/\mu^2)$ w. r.
t. the leading term.

The limit of {\it Multi-Regge Kinematics} (MRK) for the subprocess
(\ref{eqI:ax_proc}) is defined    as:
 \begin{eqnarray}
& \Delta y_1 =y(k_1)-y(P_{\cal A}) \gg 1,\ \Delta y_2=y(P_{\cal A})-y(k_2) \gg 1 , \label{eqI:MRK1} \\
& {\bf k}_{T1}^2\sim {\bf k}_{T2}^2 \sim M_{T{\cal A}}^2\sim \mu^2 \ll s, \label{eqI:MRK2}
 \end{eqnarray}
where rapidity for the four-momentum $k$ is equal to
$y(k)=\frac{1}{2}\log\left( \frac{k^+}{k^-} \right)$. The rapidity
gaps $\Delta y_{1,2}$ can be calculated as:
 \[
 \Delta y_{1,2}=\log\left[ \frac{M_{T{\cal A}}}{|{\bf k}_{T1,2}|} \frac{1-z_{1,2}}{z_{1,2} - \frac{{\bf k}_{T2,1}^2}{s(1-z_{2,1})}} \right].
 \]
From this expression, taken together with the conditions
(\ref{eqI:MRK1}) and (\ref{eqI:MRK2}), one can see, that the
following hierarchy holds in the MRK limit:
 \begin{equation}
 \frac{{\bf k}_{T1,2}^2}{s} \ll z_1\sim z_2 \ll 1, \label{eqI:MRK-hier}
 \end{equation}
so that the small parameters, which control the MRK limit are
actually $z_{1,2}$, while the transverse momenta are of the same
order of magnitude as the hard scale, and the collinear asymptotic
of the amplitude (\ref{eqI:CL_M2}) is inapplicable. Also, the
following scaling relations for momentum components hold in the MRK
limit:
 \begin{equation}
 M_{T{\cal A}}\sim |{\bf k}_{T1}|\sim q_1^+ \sim O(z_1) \gg q_1^- \sim O(z_1^2),\
  M_{T{\cal A}}\sim |{\bf k}_{T2}|\sim q_2^- \sim O(z_2) \gg q_2^+ \sim O(z_2^2), \label{eqI:scaling}
 \end{equation}
which allows one to neglet the ``small'' light-cone components of
momenta $q_1^-$ and $q_2^+$.

The systematic formalism for the calculation of the asymptotic
expressions for arbitrary QCD amplitudes in the MRK limit has been
formulated by L. N. Lipatov and M. I. Vyazovsky in a form of
gauge-invariant Effective Field Theory (EFT) for Multi-Regge
processes in QCD~\cite{Lipatov95, LipatovVyazovsky}, see
also~\cite{LipatovRev} for a review. The MRK asymptotics of the
amplitude in this EFT is constructed from gauge-invariant blocks --
{\it effective vertices}, which describe the production of clusters
of QCD partons, separated by the large rapidity gaps. These
effective vertices are connected together via $t$-channel exchanges
of gauge-invariant off-shell degrees of freedom, Reggeized gluons
$R_{\pm}$ and Reggeized quarks $Q_{\pm}$. The latter obey special
kinematical constraints, such that the field $Q_\pm$($R_\pm$)
carries only $q^\pm$ light-cone component of momentum and the
transverse momentum of the same order of magnitude, while $q^\mp=0$.
As it was shown above, these kinematical constraints are equivalent
to MRK.

\begin{figure}
 \begin{center}
 \includegraphics[width=0.7\textwidth]{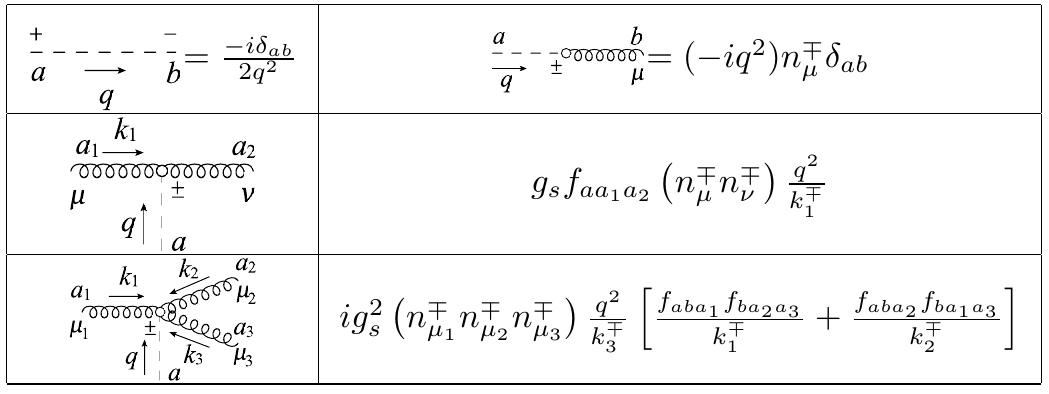}
 \end{center}
\caption{Feynman rules of the EFT~\cite{Lipatov95}. Propagator of
the Reggeized gluon (top-left panel) and Reggeon-gluon induced
vertices up to the $O(g_s^2)$ are shown. The usual Feynman Rules of
QCD hold for interactions of ordinary quarks and
gluons.\label{figI:FeynmanRules}}
\end{figure}

\begin{figure}
\begin{center}
\includegraphics[width=\textwidth]{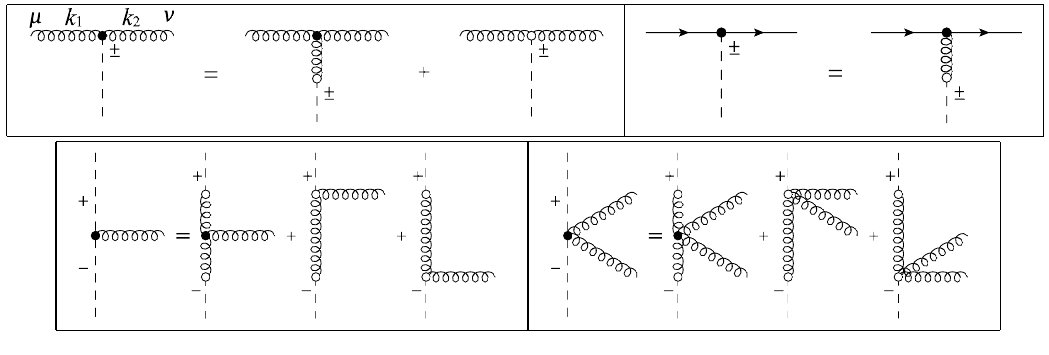}
\end{center}
\caption{Structure of the effective vertices $R_\pm gg$ (top-left),
$R_\pm q\bar{q}$ (top-right), $R_+R_- g$ (bottom-left) and the
$R_+R_-gg$ {\it combined vertex} (bottom-right). These veritces
appear in the diagrams of the Figs.~\ref{figI:MRK_ampl},
\ref{figII:RRqq} and~\ref{figII:RRqqg}.\label{figI:FeynmanRules2}}
\end{figure}

Due to the requirements of gauge-invariance of effective vertices
and the above-mentioned kinematic constraints, the interactions of
QCD partons and Reggeons in the EFT~\cite{Lipatov95,
LipatovVyazovsky} are nonlocal and contain the Wilson's exponents of
gluonic fields. After the perturbative expansion, the latter
generate an infinite series of induced vertices of interaction of
particles and Reggeons. The Feynman Rules of the EFT are worked out
in details in the Ref.~\cite{EFT_FRs}, however we also collect the
induced and effective vertices, relevant for our present study in
the Figs.~\ref{figI:FeynmanRules} and~\ref{figI:FeynmanRules2} for
the reader's convenience.

\begin{figure}
 \begin{center}
 \includegraphics[width=0.4\textwidth]{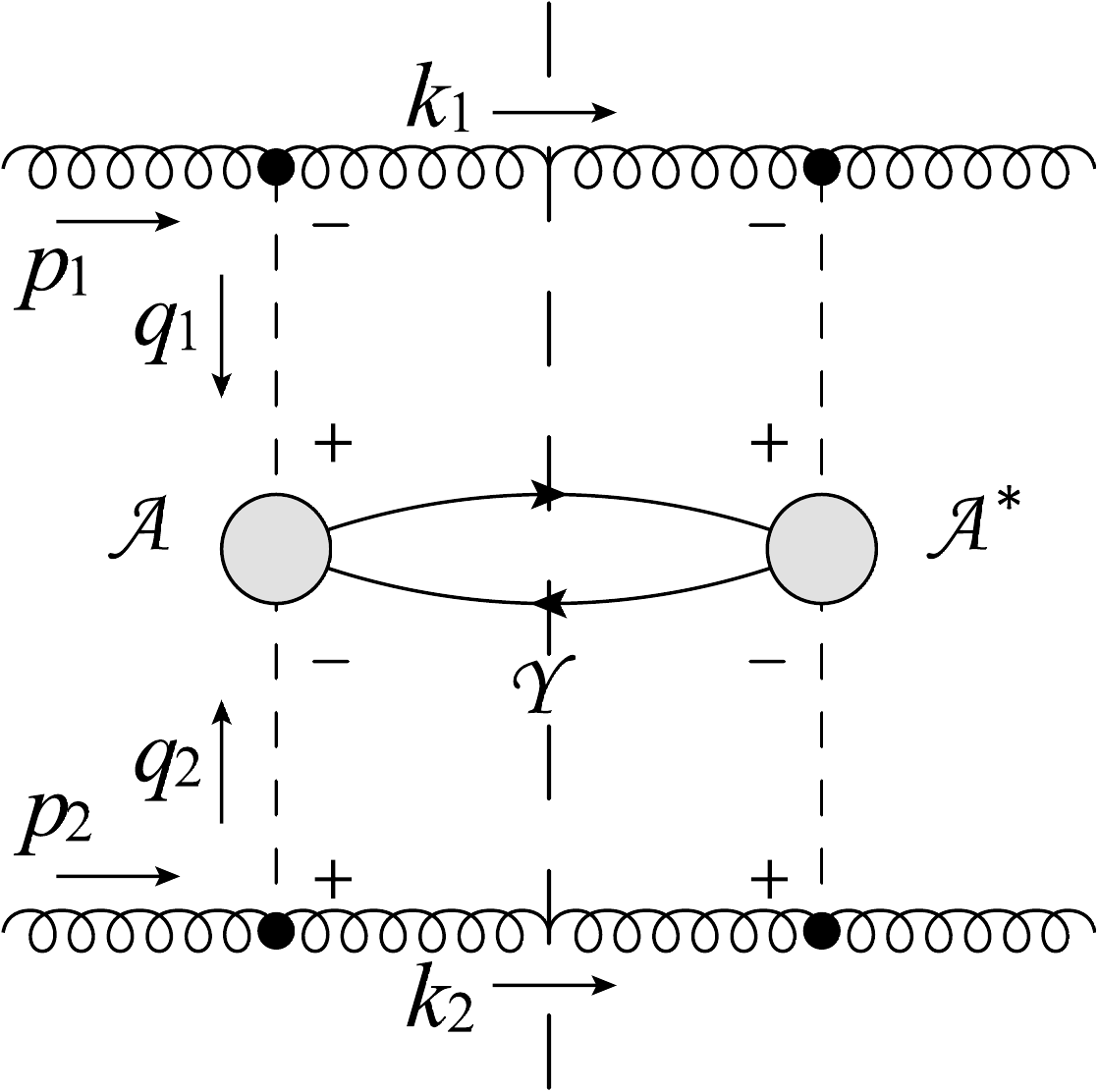}
 \end{center}
 \caption{Diagrammatic representation of the MRK asymptotics for squared amplitude of the subprocess (\ref{eqI:ax_proc}).\label{figI:MRK_ampl}}
 \end{figure}

The diagrammatic representation of the squared amplitude of the
process (\ref{eqI:ax_proc}) is shown in the
Fig.~\ref{figI:MRK_ampl}. Explicitly, the $R_\pm gg$ effective
vertex, which is depicted diagrammatically in the
Fig.~\ref{figI:FeynmanRules2}, reads:
\[
\Gamma^{abc}_{\mu\nu\pm}(k_1,k_2)=-ig_s f^{abc} \left[ 2g_{\mu\nu}k_1^\mp + (2k_2+k_1)_\mu n^\mp_\nu - (2k_1+k_2)_\nu n^\mp_\mu - \frac{(k_1+k_2)^2}{k_1^\mp} n^\mp_\mu n^\mp_\nu \right].
\]
  Evaluating the square of $R_\pm gg$ effective vertex, contracted with the polarization vectors of on-shell external gluons one obtains:
 \begin{equation}
 \sum\limits_{\lambda_1,\lambda_2} \left\vert \Gamma_{\mu\nu\pm}(k_1,-k_2) \epsilon_\mu(k_1,\lambda_1) \epsilon^\star_\nu(k_2,\lambda_2) \right\vert^2 = 8(k_1^\mp)^2. \label{eqI:Gmn_sq}
 \end{equation}

Using the result (\ref{eqI:Gmn_sq}) and the Feynman rules of the
Fig.~\ref{figI:FeynmanRules} one can write the MRK asymptotics of
the squared amplitude of the process (\ref{eqI:ax_proc}) in the
following form:
 \begin{equation}
  \overline{|{\cal M}|^2}_{\rm MRK} \simeq \frac{4g_s^4}{{\bf k}_{T1}^2 {\bf k}_{T2}^2} \tilde{P}_{gg}(z_1) \tilde{P}_{gg}(z_2) \frac{\overline{|{\cal A}_{PRA}|^2}}{z_1 z_2}, \label{eqI:MRK_M2}
 \end{equation}
where the MRK gluon-gluon splitting functions
$\tilde{P}_{gg}(z)=2C_A/z$ reproduce the small-$z$ asymptotics of
the full DGLAP splitting functions and the squared PRA amplitude is
defined as:
 \begin{equation}
 \overline{ |{\cal A}_{PRA}|^2} = \left( \frac{q_1^+q_2^-}{4(N_c^2-1)\sqrt{t_1t_2}} \right)^2 \left[ {\cal A}_{c_1 c_2}^\star {\cal A}^{c_1 c_2} \right],
 \end{equation}
where ${\cal A}$ is the Green's function of the subprocess $R_+(q_1)
+ R_-(q_2)\to {\cal Y}(P_{\cal A})$ with amputated propagators of
the Reggeized gluons, and $c_{1,2}$ are their color indices. The error of the approximation
(\ref{eqI:MRK_M2}) is supressed as $O(z_{1,2})$ w. r. t. the leading
term.

In contrast with the collinear limit, PRA amplitude explicitly and
nontrivially depends on the ${\bf q}_{T1}$ and ${\bf q}_{T2}$.
However, when ${\bf k}_{T1,2}\ll \mu^2$, MRK limit reduces to the
small-$z_{1,2}$ asymptotics of the collinear limit and the
Eq.~(\ref{eqI:MRK_M2}) should reproduce the Eq.~(\ref{eqI:CL_M2}).
To this end, the following {\it collinear limit constraint} for the
PRA amplitude should hold:
 \begin{equation}
 \int\frac{d\phi_1 d\phi_2}{(2\pi)^2} \lim\limits_{t_{1,2}\to 0}  \overline{ |{\cal A}_{PRA}|^2} = \overline{|{\cal A}_{CPM}|^2}, \label{eqI:CL_A}
 \end{equation}
where $\phi_{1,2}$ are the azimuthal angles of the vectors ${\bf
q}_{T1,2}$. One can prove the constraint
(\ref{eqI:CL_A}) for the general PRA amplitudes of the type
$R_++R_-\to {\cal Y}$, with the help of Ward identities for the
Green's functions with Reggeized gluons, which has been discovered
in the Ref.~\cite{BLV}.

Now we introduce the {\it modified MRK (mMRK) approximation} for the
squared amplitude of the subprocess (\ref{eqI:ax_proc}) as follows:
 \begin{enumerate}
 \item In the Eq. (\ref{eqI:MRK_M2}) we substitute the MRK asymptotics for the splitting fuctions $\tilde{P}_{gg}(z)$ by the full LO DGLAP expression $P_{gg}(z)$.
 \item We substitute the factors ${\bf k}_{T1,2}^2$ in the denominator of (\ref{eqI:MRK_M2}) by the exact value of $q^2_{1,2}$, as if all four components of momentum $q_{1,2}^+$, $q_{1,2}^-$ and ${\bf q}_T$ where flowing through the $t$-channel propagator: ${\bf k}_{T1,2}^2\to -q_{1,2}^2={\bf q}_{T1,2}^2/(1-z_{1,2})$.
 \item However, the ``small'' light-cone components of momenta: $q_1^-$ and $q_2^+$ do not propagate into the hard scattering process, so it's gauge-invariant definition is unaffected and is given by the Lipatov's EFT~\cite{Lipatov95}.
 \end{enumerate}
After these substitutions, the mMRK approximation for the squared amplitude of the subprocess (\ref{eqI:ax_proc}) takes the following form:
\begin{equation}
  \overline{|{\cal M}|^2}_{\rm mMRK} \simeq \frac{4g_s^4}{q_{1}^2 q_{2}^2} P_{gg}(z_1) P_{gg}(z_2) \frac{\overline{|{\cal A}_{PRA}|^2}}{z_1 z_2}. \label{eqI:mMRK_M2}
\end{equation}

The mMRK approximation (\ref{eqI:mMRK_M2}) reproduces the exact QCD
results both in the collinear and MRK limits. The latter suggests,
that it should be more accurate than the default collinear limit
approximation (\ref{eqI:CL_M2}) when ${\bf k}_{T1,2}\sim \mu^2$ even
outside of the strict MRK limit $z_{1,2}\ll 1$, however at present
we can not give the precise parametric estimate of accuracy of the
Eq.~(\ref{eqI:mMRK_M2}) in this kinematic region. The available
numerical evidence (see the Ref.~\cite{HEJ1} for the case of
amplitudes with reggeized gluons in the $t$-channel and
Refs.~\cite{HHJ, Diphotons} for the case of Reggeized quarks)
supports the form of mMRK approximation, proposed above.

To derive the LO factorization formula of PRA we substitute the mMRK
approximation (\ref{eqI:mMRK_M2}) to the factorization formula of
CPM integrated over the phase-space of additional partons $k_{1,2}$:
  \begin{eqnarray}
  d\sigma &=& \int \frac{dk_1^+ d^2{\bf k}_{T1}}{(2\pi)^3 k_1^+} \int \frac{dk_2^- d^2{\bf k}_{T2}}{(2\pi)^3 k_2^-}
  \int d\tilde{x}_1 d\tilde{x}_2 f_g(\tilde{x}_1,\mu^2) f_g(\tilde{x}_2,\mu^2)\ \frac{\overline{|{\cal M}|^2}_{\rm mMRK}}{2S\tilde{x}_1\tilde{x}_2}
  \times \nonumber \\
&\times & (2\pi)^4 \delta\left( \frac{1}{2}\left( q_1^+ n_- + q_2^-
n_+ \right) + q_{T1}+ q_{T2} - P_{\cal A} \right)  d\Phi_{\cal A},
\label{eqI:dsig0}
  \end{eqnarray}
where $f_g(x,\mu^2)$ are the (integrated) Parton Distribution
Functions (PDFs) of the CPM,
$p_{1,2}^\mu=\tilde{x}_{1,2}P_{1,2}^\mu$, where $P_{1,2}$ are the
four-momenta of colliding protons, and $d\Phi_{\cal A}$ is the
element of the Lorentz-invariant phase-space for the final state of
the hard subprocess ${\cal Y}$.

Changing the variables in the integral: $(k_1^+, \tilde{x}_1)\to
(z_1, x_1)$, $(k_2^-,\tilde{x}_2)\to (z_2,x_2)$, where
$x_{1,2}=\tilde{x}_{1,2}z_{1,2}$, one can rewrite the
Eq.~\ref{eqI:dsig0} in a $k_T$-factorized form:
  \begin{eqnarray}
  d\sigma &=& \int\limits_0^1 \frac{dx_1}{x_1} \int \frac{d^2{\bf q}_{T1}}{\pi} \tilde{\Phi}_g(x_1,t_1,\mu^2) \int\limits_0^1 \frac{dx_2}{x_2} \int \frac{d^2{\bf q}_{T2}}{\pi} \tilde{\Phi}_g(x_2,t_2,\mu^2)\cdot d\hat{\sigma}_{\rm PRA}, \label{eqI:kT_fact}
  \end{eqnarray}
where the partonic cross-section in PRA is given by:
  \begin{equation}
  d\hat{\sigma}_{\rm PRA}= \frac{\overline{|{\cal A}_{PRA}|^2}}{2Sx_1x_2}\cdot (2\pi)^4 \delta\left( \frac{1}{2}\left( q_1^+ n_- + q_2^- n_+ \right) + q_{T1}+ q_{T2} - P_{\cal A} \right)  d\Phi_{\cal A}, \label{eqI:CS_PRA}
  \end{equation}
and the tree-level ``unintegrated PDFs'' (unPDFs) are:
  \begin{equation}
  \tilde{\Phi}_g(x,t,\mu^2) = \frac{1}{t} \frac{\alpha_s}{2\pi} \int\limits_x^1 dz\ P_{gg}(z)\cdot \frac{x}{z}f_g\left(\frac{x}{z},\mu^2\right). \label{eqI:tree_unPDFs}
  \end{equation}

The cross-section (\ref{eqI:kT_fact}) with ``unPDFs''
(\ref{eqI:tree_unPDFs}) contains the collinear divergence at
$t_{1,2}\to 0$ and infrared (IR) divergence at $z_{1,2}\to 1$. To
regularize the latter, we observe, that the mMRK expression
(\ref{eqI:mMRK_M2}) can be expected to give a reasonable
approximation for the exact matrix element only in the
rapidity-ordered part of the phase-space, where $\Delta y_1>0$ and
$\Delta y_2>0$. The cutoff on $z_{1,2}$ follows from this
conditions:
  \begin{equation}
  z_{1,2}< 1-\Delta_{KMR}(t_{1,2},\mu^2), \label{eq:KMR_cut}
  \end{equation}
  where $\Delta_{KMR}(t,\mu^2)=\sqrt{t}/(\sqrt{\mu^2}+\sqrt{t})$, and we have taken into account that $\mu^2\sim M_{T{\cal A}}^2$. The collinear singularity is regularized by the Sudakov formfactor:
  \begin{equation}
  T_i(t,\mu^2)=\exp\left[ - \int\limits_t^{\mu^2} \frac{dt'}{t'} \frac{\alpha_s(t')}{2\pi} \sum\limits_{j=q,\bar{q},g} \int\limits_0^{1} dz\ z\cdot P_{ji}(z) \theta\left(1-\Delta_{KMR}(t',\mu^2) - z\right)  \right], \label{eq:Sudakov}
  \end{equation}
  which resums doubly-logarithmic corrections $\sim\log^2 (t/\mu^2)$ in the LLA in a way similar to what is done in the standard PS~\cite{PS_rev}.

The final form of our unPDF is:
  \begin{equation}
  \Phi_i(x,t,\mu^2) = \frac{T_i(t,\mu^2)}{t} \frac{\alpha_s(t)}{2\pi} \sum_{j=q,\bar{q},g} \int\limits_x^{1} dz\ P_{ij}(z)\cdot \frac{x}{z}f_{j}\left(\frac{x}{z},t \right)\cdot \theta\left(1-\Delta_{KMR}(t,\mu^2)-z \right), \label{eqI:KMR}
  \end{equation}
which coincides with Kimber, Martin and Ryskin (KMR)
unPDF~\cite{KMR}. The KMR unPDF is actively used in the
phenomenological studies employing $k_T$-factorization, but to our
knowledge, the reasoning above is the first systematic attempt to
uncover it's relationships with MRK limit of the QCD amplitudes.

The KMR unPDF approximately (see Sec.~2 of the Ref.~\cite{NLO_KMR}
for the further details) satisfies the following normalization
condition:
  \begin{equation}
  \int\limits_0^{\mu^2} dt\ \Phi_i(x,t,\mu^2) = xf_i(x,\mu^2), \label{eqI:KMR_norm}
  \end{equation}
which ensures the normalization for the single-scale observables,
such as proton structure functions or $d\sigma/dQ^2dy$ cross-section
in the Drell-Yan process, on the corresponding LO CPM results up to
power-supressed corrections and terms of the NLO in $\alpha_s$.
Results for multiscale observables in PRA are significantly
different than in CPM, due to the nonzero transverse-momenta of
partons in the initial state.

The main difference of PRA from the multitude of studies in the
$k_T$-factorization, such as Ref.~\cite{JZL_kT-fact}, is the
application of matrix elements with off-shell initial-state partons
(Reggeized quarks and gluons) from Lipatov's EFT~\cite{Lipatov95,
LipatovVyazovsky}, which allows one to study the arbitrary processes
involving non-Abelian structure of QCD without violation of
gauge-invaiance due to the nonzero virtuality of initial-state
partons. This approach, together with KMR unPDF gives stable and
consistent results in a wide range of phenomenological applications,
which include the description of the angular correlations of
dijets~\cite{NSSjj}, $b$-jets~\cite{SSbb}, charmed~\cite{PLB2016,
PRD2016DD} and bottom-flavoured~\cite{KNSS2015}  mesons, different
multiscale observables in hadroproduction of
diphotons~\cite{Diphotons} and photoproduction of photon+jet
pairs~\cite{photon_jet}, as well as some other examples.

Recently, the new approach to derive gauge-invariant scattering
amplitudes with off-shell initial-state partons, using the
spinor-helicity techniques and BCFW-like recursion relations for
such amplitudes has been introduced in the Refs.~\cite{hameren3,
katie}. This formalism is equivalent to the Lipatov's EFT at the
tree level, but for some observables, e. g. related with heavy
quarkonia, or for the generalization of the formalism to NLO, the
explicit Feynman rules and the structure of EFT is more convenient.

\section{LO PRA merged with tree-level NLO corrections}
\label{sec:merging}

Let's consider the kinematic conditions of a measurement in the
Ref.~\cite{bCMS}. In this experiment, the events with at least one
jet having $p_T^{\rm jet}>p_{TL}^{\min}$ has been recorded in
$pp$-collisions at the $\sqrt{S}=7$ TeV, and the
semileptonic decays of $B$-hadrons where reconstructed in this
events, through the decay vertices, displaced w.~r.~t. the primary
$pp$-collision vertex. The $B$-hadron is required to have
$p_{TB}>p_{TB}^{\min}=15$ GeV, while three data-samples are
presented in the Ref.~\cite{bCMS} for three values of
$p_{TL}^{\min}=56$, $84$ and $120$ GeV. The rapidities of
$B$-hadrons are constrained to be $|y_B|<y_B^{\max}=2$, while the
leading jet is searched in somewhat wider domain $|y_{\rm
jet}|<y_{\rm jet}^{\max}=3$.

The leading jet, reconstructed in this experiment, sets the hard
scale of the event. Two possibilities should be considered: the
first one is, that the jet originating from $b$-quark or
$\bar{b}$-antiquark is the leading one, and the second option is,
that some gluon or light-quark jet is leading in $p_T$, and jets
originating from $b$ or $\bar{b}$ are subleading. Observables with
such kinematic constraints on the QCD radiation are difficult to
study in $k_T$-factorization, because the radiation of additional
hard partons is already taken into account in the unPDFs, and the
jet, originating from unPDF could happen to be the leading one.

\begin{figure}
\begin{center}
\includegraphics[width=0.55\textwidth]{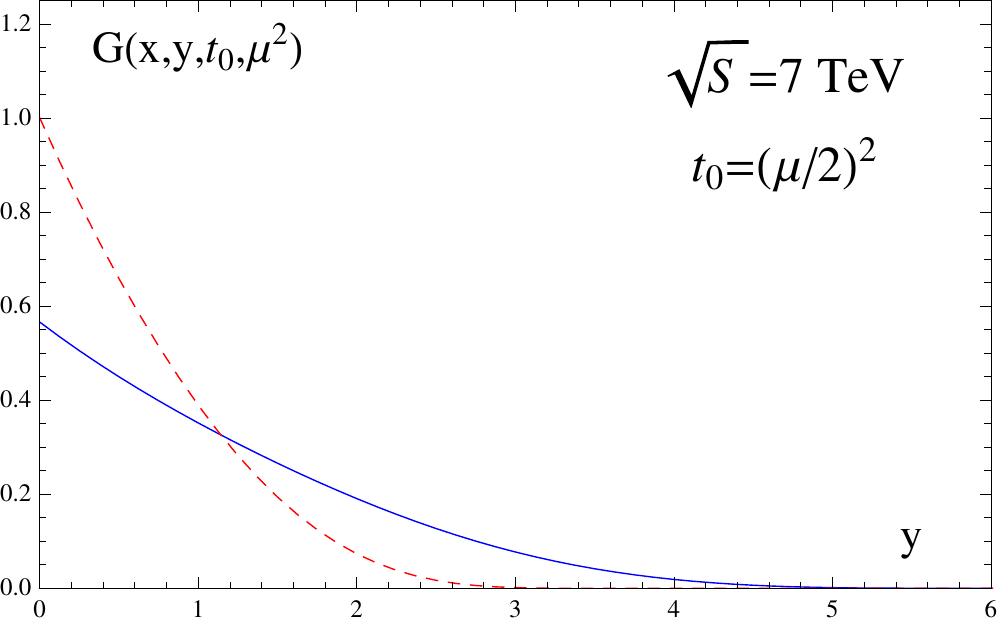}
\end{center}
\caption{ Distribution in rapidity of a gluon jets with $|{\bf
k}_T|>\mu/2$ from the last stage of the parton cascade, as given by
the KMR model~(\ref{eqI:KMR}). Solid line -- $\mu^2=10^3$ GeV$^2$,
dashed line -- $\mu^2=10^5$ GeV$^2$. Both plots are nomalized to the
common integral, scale of the $G$-axis is arbitrary. For both
distributions: $x=\mu/\sqrt{S}$, i. e. the rapidity of the hard
process is zero. \label{figII:rapidity_plots} }
  \end{figure}

One can easily estimate the distribution of additional jets in
rapidity, using the KMR model for unPDFs (\ref{eqI:KMR}).
The variable $z$ is related with rapidity ($y$) of the parton,
emitted on the last step of the parton cascade, as follows:
  \[
  z(y)= \left(1+\frac{\sqrt{t}}{x\sqrt{S}}e^{y}\right)^{-1},
  \]
so, starting from Eq.~(\ref{eqI:KMR}) one can derive the
distribution integrated over $t$ from some scale $t_0$ up to
$\mu^2$, but unintegrated over $y$: $G_i(x,y,t_0,\mu^2)$.
Representative plots of this distribution for the case of
$P_{gg}$-splitting only, are shown in the Fig.~\ref{figII:rapidity_plots}
for some values of scales typical for the process under
consideration. The LO PDFs from the
MSTW-2008 set~\cite{MSTW-2008} has been used to produce this plot. 

 From Fig.~\ref{figII:rapidity_plots} it is clear,
that in the KMR model, the majority of hardest ISR jets jets with
${\bf k}_T^2\sim \mu^2$ lie within the rapidity interval $|y|<3$  if
the particles, produced in the primary hard process have rapidities
close to zero. Therefore this jets can be identified as the leading
ones. But the kinematic approximations, which has been made in the
derivation of the factorization formula, are least reliable in this
region of phase-space, and hence the poor agreement with data is to
be expected. To avoid the above-mentioned problem, we will merge the
LO PRA description of events with the leading $b$($\bar{b}$)-jets
with events triggered by the leading gluon jet, originating from the
exact $2\to 3$ NLO PRA matrix element.

The LO ($O(\alpha_s^2)$) subprocess, which we will take into account
is:
   \begin{equation}
   R_+(q_1) + R_-(q_2)\to b(q_3)(\rightarrow B (p_{TB})) + \bar{b}(q_4)(\rightarrow \bar{B} (p_{T\bar{B}})  ),\label{eqII:LO_subpr}
   \end{equation}
where the hadronization of $b$($\bar{b}$)-quarks into the
$B$($\bar{B}$) mesons is described by the set of universal,
scale-dependent parton-to-hadron fragmentation functions, fitted on
the world data on the $B$-hadron production in $e^+e^-$-annihilation
in the Ref.~\cite{FFB}.

The following kinematic cuts are applied to the LO subprocess
(\ref{eqII:LO_subpr}):
   \begin{enumerate}
\item Both $B$ and $\bar{B}$ mesons are required to have $|y_B|<y_B^{\max}$ and $\min(p_{TB},p_{T\bar{B}})> p_{TB}^{\min}$.
\item If the distance between three-momenta ${\bf q}_3$ and ${\bf q}_4$ in the $(\Delta y, \Delta \phi)$-plane:
$\Delta R_{34}=\sqrt{\Delta y_{34}^2+\Delta\phi_{34}^2}>\Delta R_{\rm exp.}=0.5$, then $b$ and $\bar{b}$ jets
are resolved separately and we define: $p_{TL}=\max(|{\bf q}_{T3}|,|{\bf q}_{T4}|)$.
\item If $\Delta R_{34}<\Delta R_{\rm exp.}$, then $p_{TL}=|{\bf q}_{T3}+{\bf q}_{T4}|$,
according to the anti-$k_T$ jet clustering algorithm~\cite{antikt}.
\item The MC event is {\it accepted} if $\max(|{\bf q}_{T1}|,|{\bf q}_{T2}| ) < p_{TL}$ and $p_{TL}>p_{TL}^{\min}$.
\end{enumerate}

\begin{figure}
\begin{center}
\includegraphics[width=.8\textwidth, clip=]{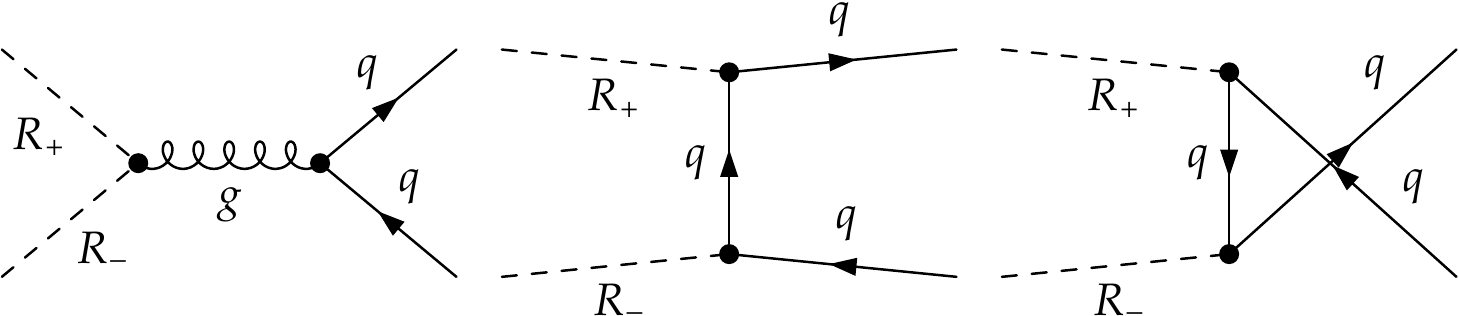}
\caption{The Feynman diagrams of Lipatov's EFT for the Reggeized
amplitude of subprocess $R_++R_-\to b+\bar b$. \label{figII:RRqq}}
\end{center}
\end{figure}

The set of Feynman diagrams for the subprocess (\ref{eqII:LO_subpr})
is presented in the Fig.~\ref{figII:RRqq}. The convinient expression
for the squared amplitude of this subprocess with massless quarks
can be found in the Ref.~\cite{NSSjj}. Due to the Ward identities of
the Ref.~\cite{BLV}, this amplitude coincides with the amplitude,
which can be obtained in the ``old $k_T$-factorization''
prescription, i. e. by substituting the polarization vectors of
initial-state gluons in the usual $gg\to q\bar{q}$ amplitude by
$q_{T1,2}^\mu/|{\bf q}_{T1,2}|$.

The NLO ($O(\alpha_s^3)$) subprocess is
   \begin{equation}
   R_+(q_1)+R_-(q_2)\to b(q_3)(\rightarrow B (p_{TB})) + \bar{b}(q_4)(\rightarrow \bar{B} (p_{T\bar{B}})) + g(q_5), \label{eqII:NLO_subpr}
   \end{equation}
and the following kinematic constraints are applied in the
calculation of this contribution:
   \begin{enumerate}
   \item Both $B$ and $\bar{B}$ mesons are required to have $|y_B|<y_B^{\max}$ and $\min(p_{TB},p_{T\bar{B}})> p_{TB}^{\min}$.
   \item Gluon jet is the leading one: $p_{TL}=|{\bf q}_{T5}|$, $\max\left( |{\bf q}_{T1}|, |{\bf q}_{T2}|, |{\bf q}_{T3}|, |{\bf q}_{T4}|  \right) < p_{TL}$ and $p_{TL}>p_{TL}^{\min}$.
   \item Rapidity of the gluon is required to be $|y_5|<y_{\rm jet}^{\max}$. The gluon jet is isolated: $\Delta R_{35}>\Delta R_{\rm exp.}$ and $\Delta R_{45}>\Delta R_{\rm exp.}$.
   \end{enumerate}

Furthermore, since the matrix elements for both subprocesses
(\ref{eqII:LO_subpr}) and (\ref{eqII:NLO_subpr}) are taken in the
approximation of massless $b$-quarks, the corresponding final-state
collinear singularity is regularized by the condition
$(q_3+q_4)^2>4m_b^2$, where $m_b=4.5$ GeV.

Few comments are in order. For the both subprocesses
(\ref{eqII:LO_subpr}) and (\ref{eqII:NLO_subpr}), transverse momenta
of jets from the unPDFs are constrained to be subleading. In such a
way we avoid the double-counting of the leading emissions between LO
and NLO contributions and additional subtractions are not needed.
This is in contrast to the observables fully inclusive in the QCD
radiation~\cite{Diphotons}, where the double-counting subtractions
between LO and NLO terms has to be done. Another comment concerns
the isolation condition for the leading gluon jet in the NLO
contribution (\ref{eqII:NLO_subpr}). This condition regularizes the
collinear singularity between the final-state gluon and
$b$($\bar{b}$)-quark. In the full NLO calculation, this singularity
will be cancelled by the loop correction, producing some finite
contribution, but since the gluon is required to be harder than $b$
or $\bar{b}$-quarks, this finite contribution will be proportional
to the $P_{gq}(z)=C_F(1+(1-z)^2)/z$ splitting function at $z\to 1$,
so we don't expect the logarithmically-enchanced contributions from
this region of the phase-space.

\begin{figure}
\begin{center}
\includegraphics[width=.8\textwidth, clip=]{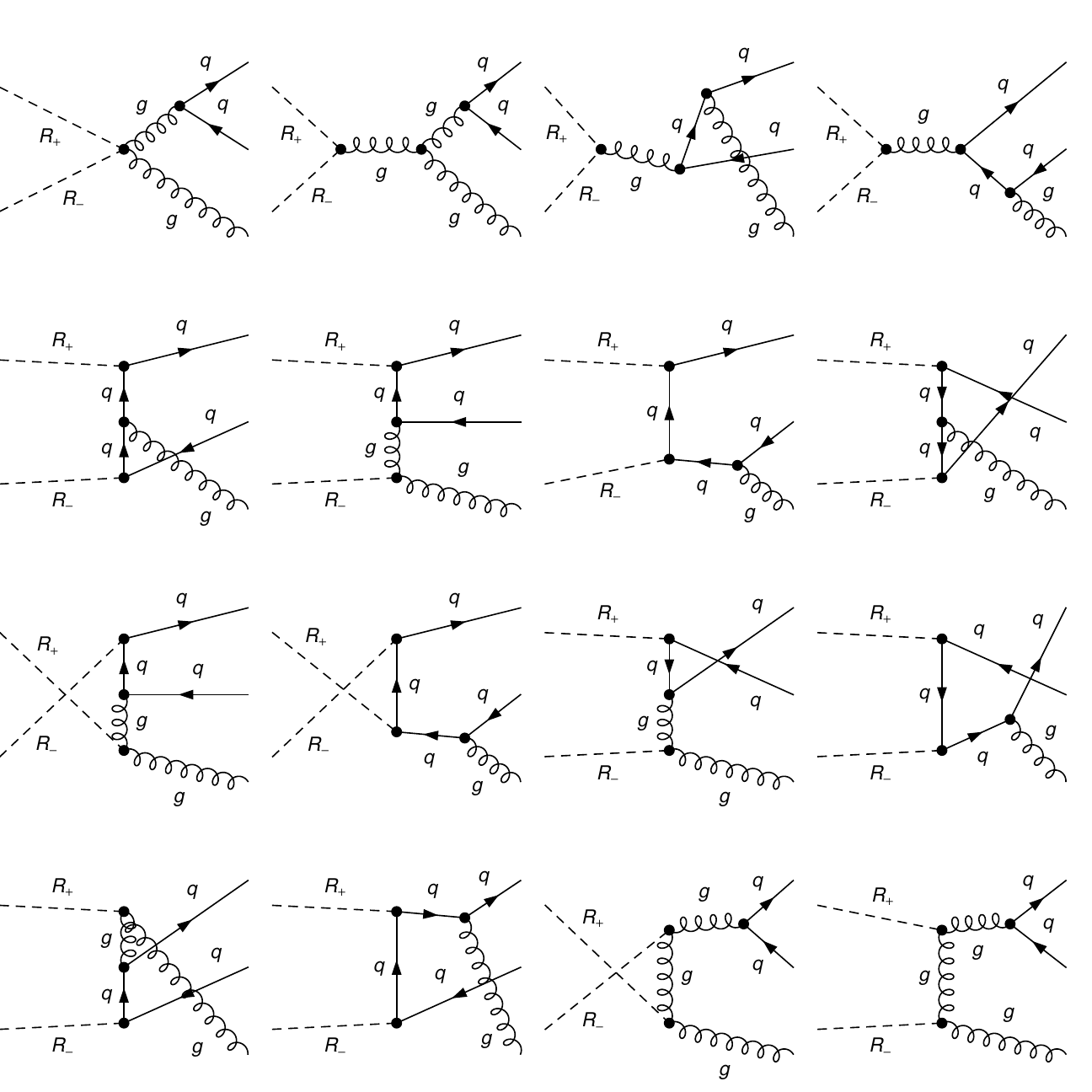}
\caption{The Feynman diagrams of Lipatov's EFT for the Reggeized
amplitude of subprocess $R_++R_-\to b+\bar b+ g$.
\label{figII:RRqqg}}
\end{center}
\end{figure}

The set of Feynman diagrams for amplitude of the subprocess
(\ref{eqII:NLO_subpr}) is presented in the Fig.~\ref{figII:RRqqg}.
We generate this amplitude, using our model-file \texttt{ReggeQCD},
which implements the Feynman rules of Lipatov's EFT in
\texttt{FeynArts}~\cite{FeynArts}. The squared ampitude is computed
using the \texttt{FormCalc}~\cite{FormCalc} package, and has been
compared numerically with the squared amplitude, obtained by the
methods of the Refs.~\cite{hameren3, katie}. Our results and
results of Ref. \cite{hameren3, katie} agree up to machine precision. Apart
from the Feynman rules, depicted in the
Figs.~\ref{figI:FeynmanRules} and \ref{figI:FeynmanRules2},
\texttt{ReggeQCD} package contains all Feynman rules, which are
needed to generate arbitrary PRA amplitude with reggeized gluons or
quarks in the initial state and up to three quarks, gluons or
photons in the final state. We are planning to publish the \texttt{ReggeQCD} model-file in a separate paper~\cite{ReggeQCD}. The \texttt{FORTRAN} code for the squared amplitudes of the processes (\ref{eqII:LO_subpr}) and (\ref{eqII:NLO_subpr}) is available from authors by request.

As it was stated above, we will use the fragmentation model, to
describe hadronization of $b$-quarks into $B$-hadrons, so the
observable cross-section is:
   \begin{eqnarray}
   \frac{d\sigma_{\rm obs.}}{dy_B dy_{\bar{B}} d\Delta\phi} &=& \int\limits_{p_{TB}^{\min}}^\infty dp_{TB} \int\limits_{p_{TB}^{\min}}^\infty dp_{T\bar{B}} \int\limits_0^1 \frac{dz_1}{z_1} D_{B/b}(z_1,\mu^2) \int\limits_0^1 \frac{dz_2}{z_2} D_{B/b}(z_2,\mu^2) \nonumber \\
   &\times & \frac{d\sigma_{b\bar{b}}}{dq_{T3} dq_{T4} dy_3 dy_{4} d\Delta\phi}, \label{eqII:frag}
   \end{eqnarray}
   where $\Delta\phi=\Delta\phi_{34}$, $D_{B/b}(z,\mu^2)$ are the fragmentation functions~\cite{FFB}, and $q_{T3}=|{\bf q}_{T3}|=p_{TB}/z_1$, $q_{T4}=|{\bf q}_{T4}|=p_{T\bar{B}}/z_2$, $y_3=y_B$, $y_4=y_{\bar B}$. To simplify the numerical calculations, it is very convenient to integrate over $q_{T3,4}$ instead of $p_{TB}$ and $p_{T\bar{B}}$ in Eq.~(\ref{eqII:frag}), then all the dependence of the cross-section on fragmentation functions can be absorbed into the following measurement function:
   \begin{equation}
   \Theta(\tilde{z},\mu^2)=\left\{ \begin{array}{c}
   \int\limits_{1/\tilde{z}}^1 dz\ D_{B/q}(z,\mu^2) \ {\rm if} \ \tilde{z}>1, \\
   0 \ {\rm otherwise,}
    \end{array}\right. \label{eqII:MF-def}
   \end{equation}
which can be efficiently computed and tabulated in advance,
therefore reducing the dimension of phase-space integrals by 2. Then
the master-formula for the cross-section of $2\to 2$ subprocess
(\ref{eqII:LO_subpr}) in PRA takes the form:

\begin{eqnarray}
    \frac{d\sigma_{\rm obs.}^{(2\to 2)}}{dy_B dy_{\bar{B}} d\Delta\phi} = \int\limits_0^\infty dq_{T3}\ dq_{T4}\cdot \Theta \left(\frac{q_{T3}}{p_{TB}^{\min}}, \mu^2 \right) \Theta \left(\frac{q_{T4}}{p_{TB}^{\min}}, \mu^2 \right)  \nonumber \\
 \times \int\limits_0^\infty dt_1 \int\limits_0^{2\pi}d\phi_1\  \Phi_g(x_1,t_1,\mu^2)  \Phi_g(x_2,t_2,\mu^2)\cdot \frac{q_{T3}q_{T4}\  \overline{\left\vert {\cal A}^{(2\to 2)}_{PRA} \right\vert^2}}{2(2\pi)^3 (Sx_1x_2)^2}\cdot \theta_{\rm cuts}^{(2\to 2)},
   \end{eqnarray}
where $\phi_1$ is azimuthal angle between the vector ${\bf q}_{T1}$
and ${\bf q}_{T3}$, $t_2=|{\bf q}_{T3}+{\bf q}_{T4}-{\bf q}_{T1}|$,
$x_1=(q_3^++q_4^+)/\sqrt{S}$, $x_2=(q_3^-+q_4^-)/\sqrt{S}$, and the
theta-function $\theta_{\rm cuts}^{(2\to 2)}$ implements the
kinematic constraints for $2\to 2$ process, described above.
Analogously, the formula for differential cross-section of the $2\to
3$ process (\ref{eqII:NLO_subpr}) reads:
    \begin{eqnarray}
    \frac{d\sigma_{\rm obs.}^{(2\to 3)}}{dy_B dy_{\bar{B}} d\Delta\phi\ dy_5 } = \int\limits_0^\infty dq_{T3}\ dq_{T4}\cdot \Theta \left(\frac{q_{T3}}{p_{TB}^{\min}}, \mu^2 \right) \Theta \left(\frac{q_{T4}}{p_{TB}^{\min}}, \mu^2 \right)  \nonumber \\
  \times  \int\limits_0^\infty dt_1 dt_2 \int\limits_0^{2\pi}d\phi_1 d\phi_2\  \Phi_g(x_1,t_1,\mu^2)  \Phi_g(x_2,t_2,\mu^2)\cdot \frac{q_{T3}q_{T4}\  \overline{\left\vert {\cal A}^{(2\to 3)}_{PRA} \right\vert^2}}{8(2\pi)^6 (Sx_1x_2)^2}\cdot \theta_{\rm cuts}^{(2\to 3)},
   \end{eqnarray}
where $\phi_2$ is the azimuthal angle between the vectors ${\bf
q}_{T2}$, and ${\bf q}_{T3}$, ${\bf q}_{T5}={\bf q}_{T1}+{\bf
q}_{T2}-{\bf q}_{T3}-{\bf q}_{T4}$,
$x_1=(q_3^++q_4^++q_5^+)/\sqrt{S}$,
$x_2=(q_3^-+q_4^-+q_5^-)/\sqrt{S}$ and the theta-function
$\theta_{\rm cuts}^{(2\to 3)}$ implements kinematic cuts for $2\to
3$ process, described after the Eq.~(\ref{eqII:NLO_subpr}).

\section{Numerical results for $B\bar{B}$-correlation observables}
\label{sec:results}

\begin{figure}[p!]
  \begin{center}
  \includegraphics[width=0.45\textwidth, angle=-90]{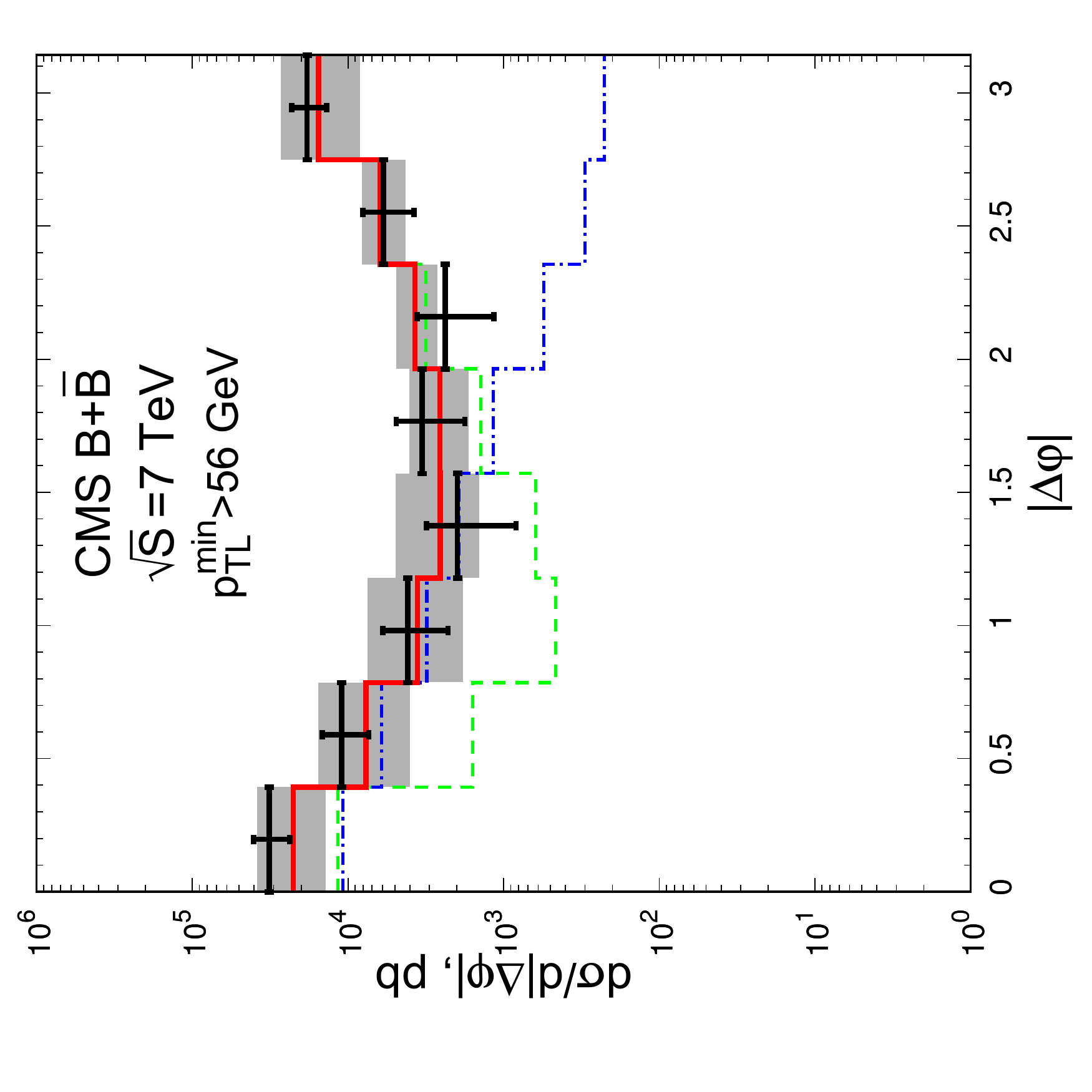}
  \includegraphics[width=0.45\textwidth, angle=-90]{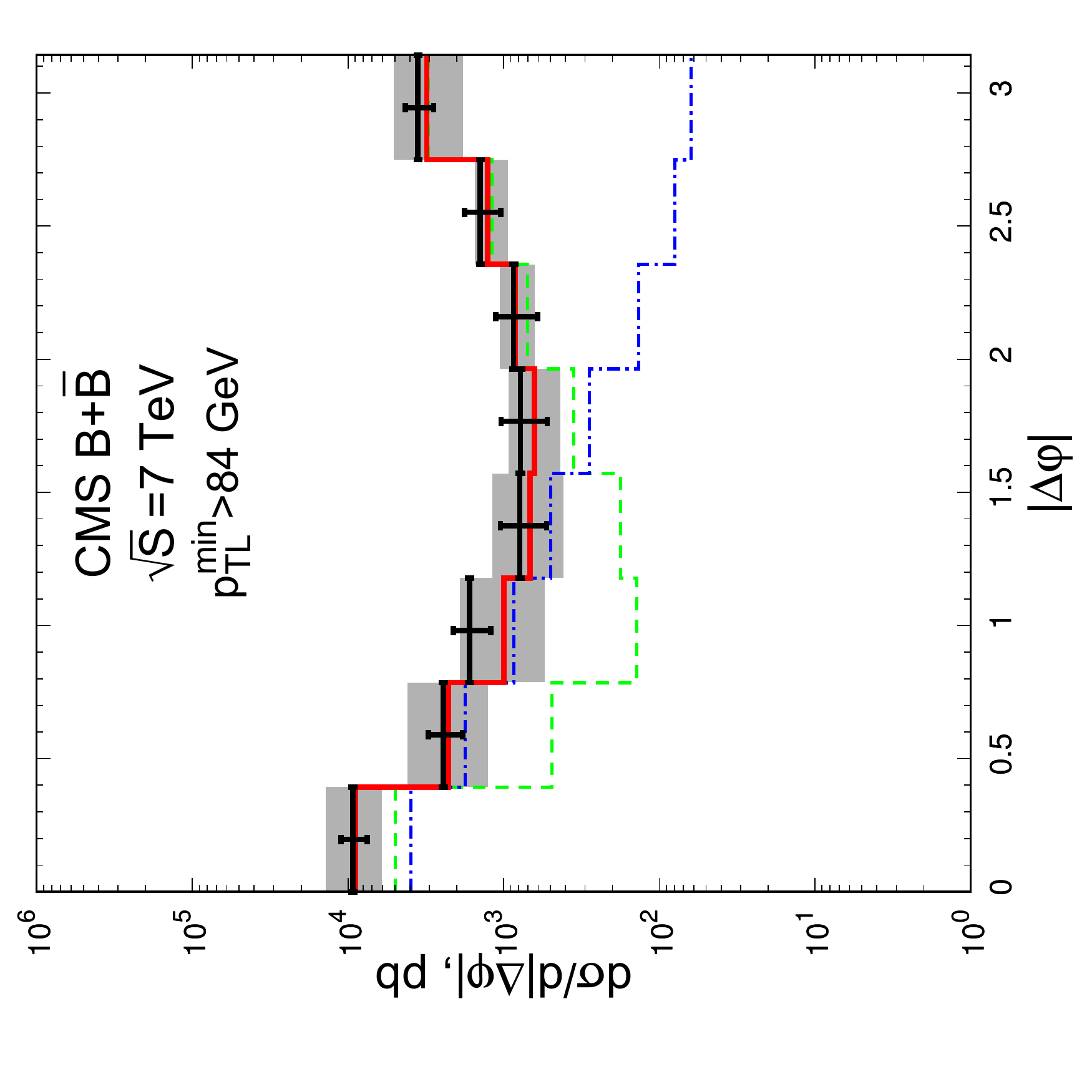} \\
  \includegraphics[width=0.45\textwidth, angle=-90]{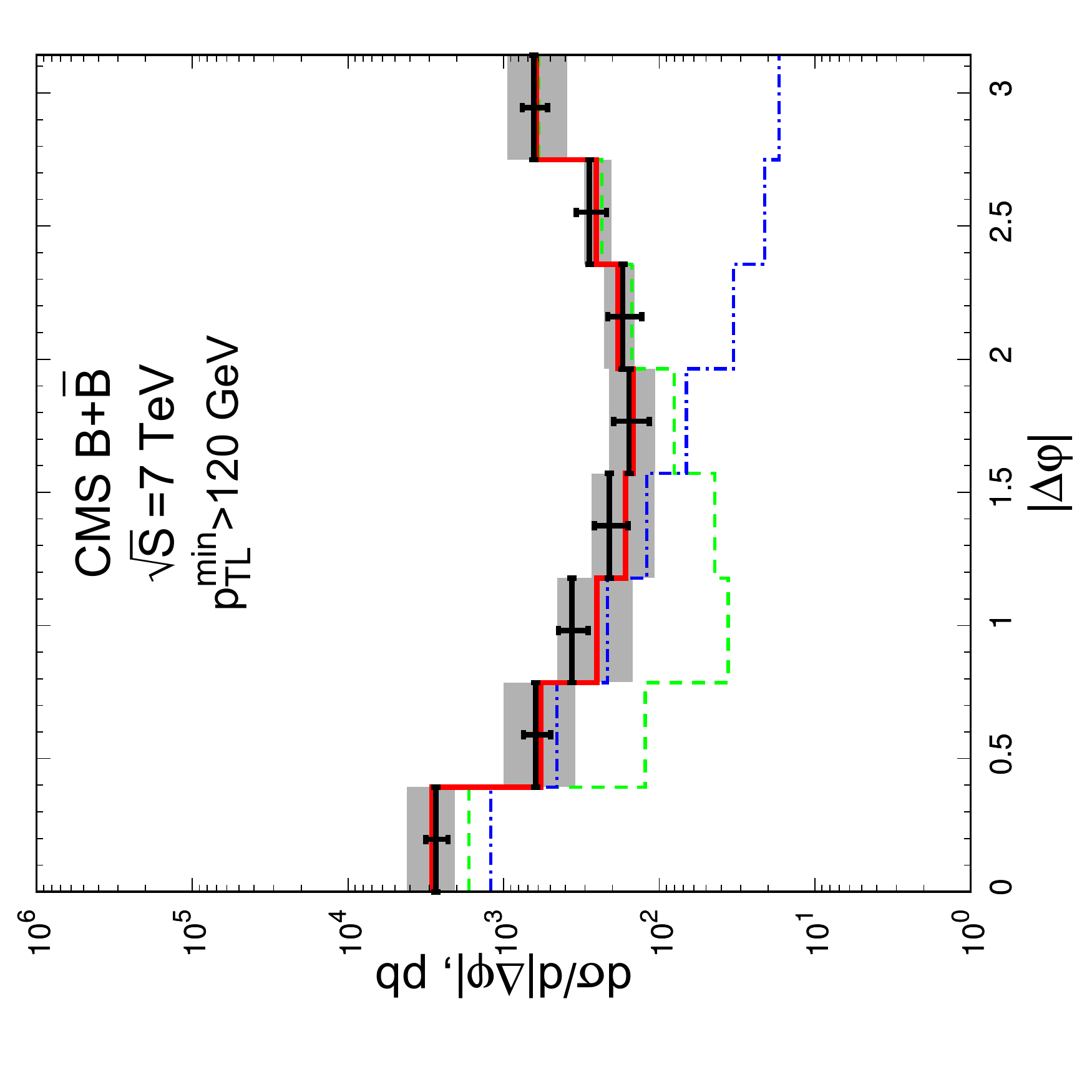}
  \end{center}
  \caption{Comparison of the predictions for $\Delta\phi$-spectra of $B\bar{B}$-pairs with the CMS data~\cite{bCMS}.
  Dashed line -- contribution of the LO subprocess (\ref{eqII:LO_subpr}), dash-dotted line -- contribution of the NLO subprocess
   (\ref{eqII:NLO_subpr}), solid line -- sum of LO and NLO contributions.  \label{figIII:dPhi-spectra}}
  \end{figure}

\begin{figure}[p!]
  \begin{center}
  \includegraphics[width=0.45\textwidth, angle=-90]{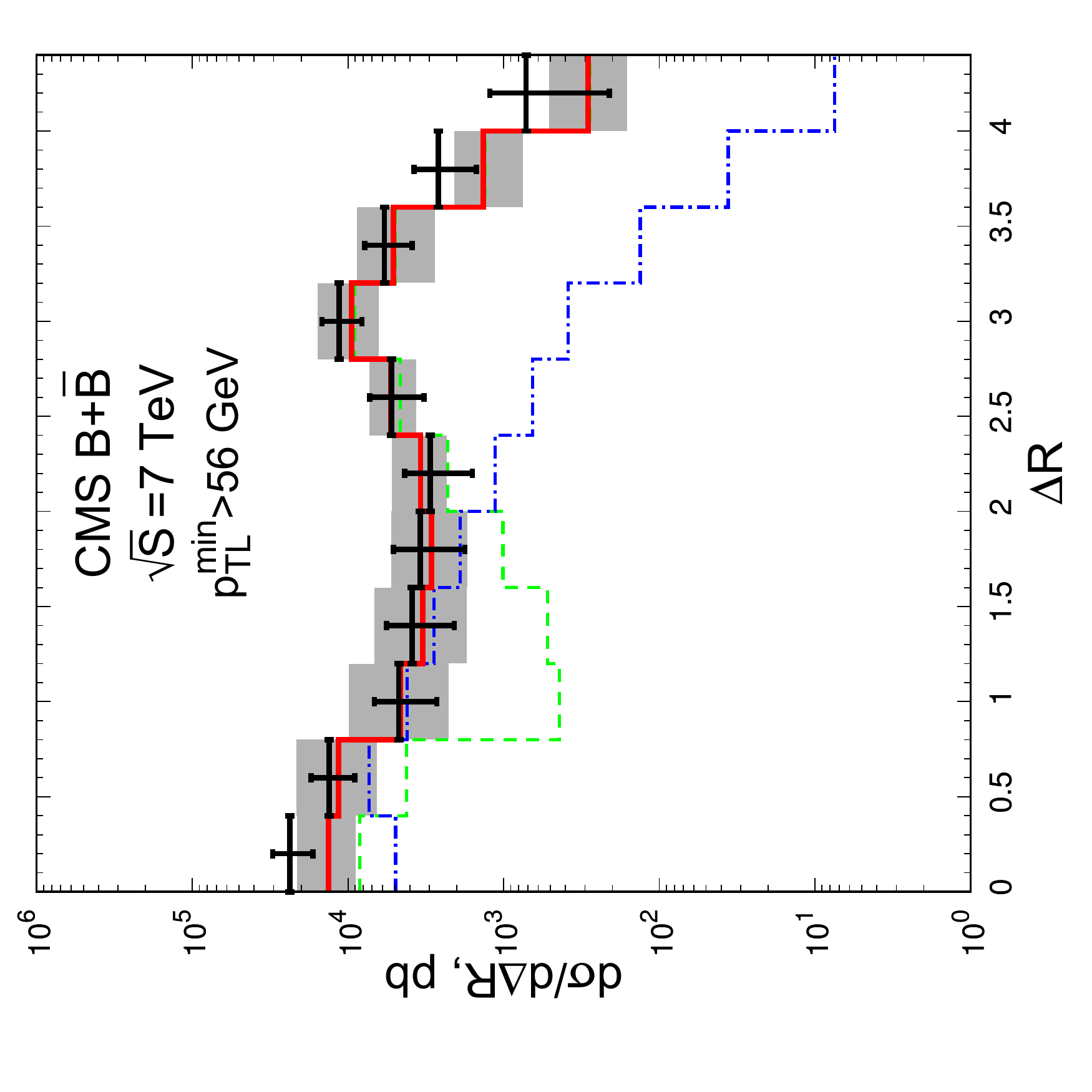}
  \includegraphics[width=0.45\textwidth, angle=-90]{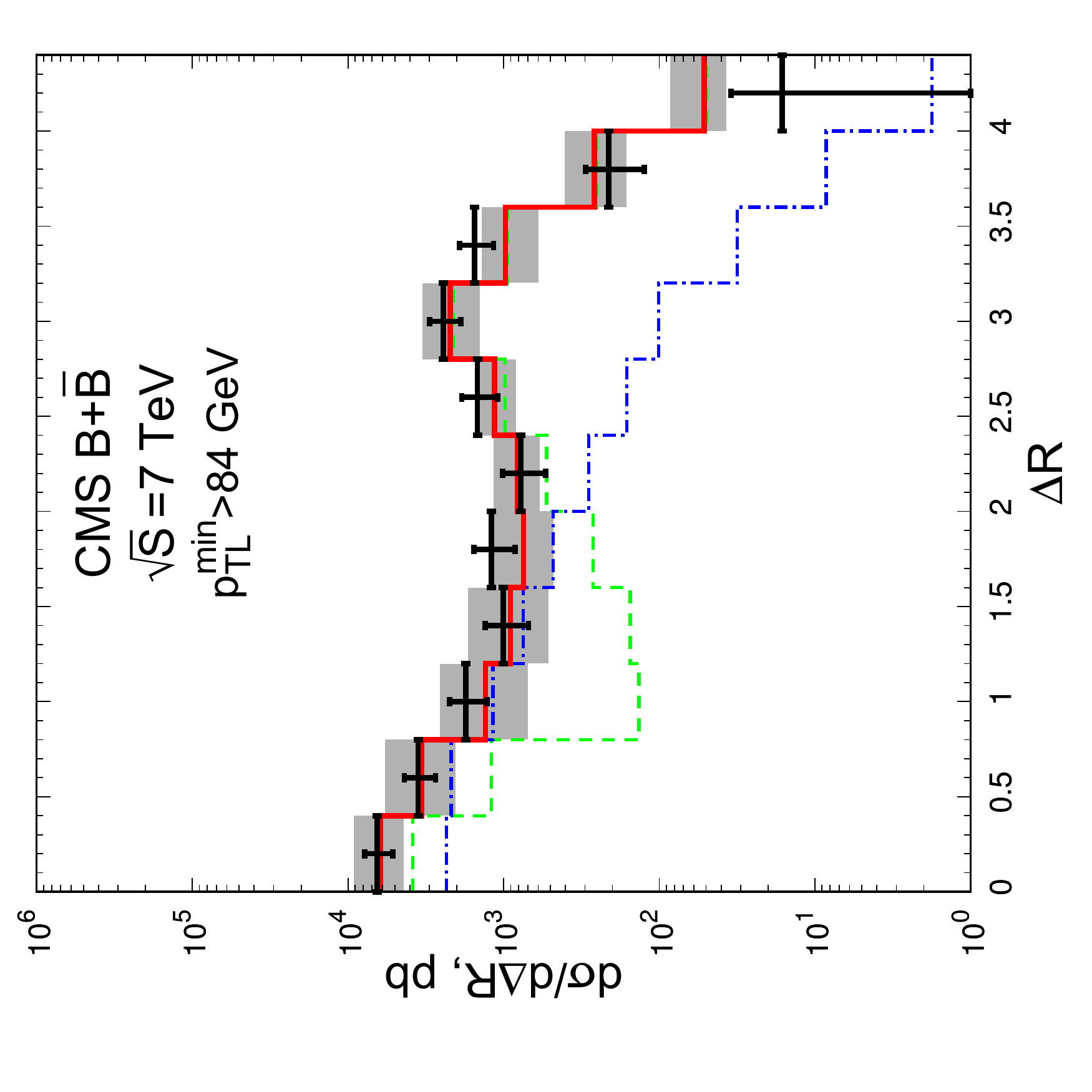} \\
  \includegraphics[width=0.45\textwidth, angle=-90]{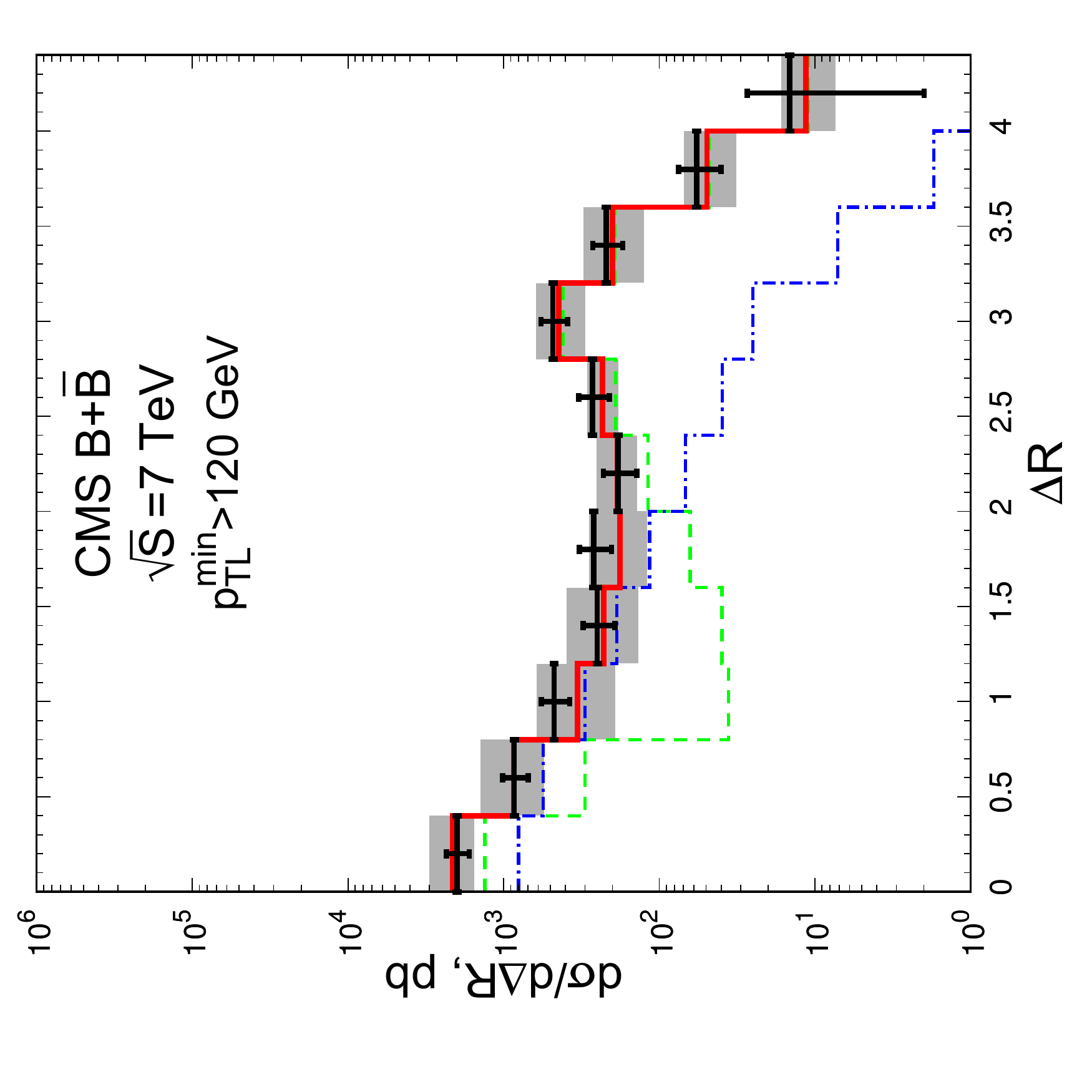}
  \end{center}
  \caption{Comparison of the predictions for $\Delta R$-spectra of $B\bar{B}$-pairs with the CMS data~\cite{bCMS}.
  Notation for the histograms is the same as in the Fig.~\ref{figIII:dPhi-spectra}. \label{figIII:dR-spectra}}
  \end{figure}

Now we are in a position to compare our numerical results, obtained
in the approximation, formulated in the Sec.~\ref{sec:merging} of
the present paper, to the experimental data of the Ref.~\cite{bCMS}.
Experimental uncertanties, related with the shape of $\Delta\phi$
and $\Delta R=\Delta R_{34}$ distributions are relatively small
($\sim 20-30\%$). They are indicated by the error-bars in the
Figs.~\ref{figIII:dPhi-spectra} and~\ref{figIII:dR-spectra}.
However, an additional uncertainty in the absolute normalization of
the cross-sections $\simeq \pm 47\%$  is reported in the
Ref.~\cite{bCMS}, and it is not included into the error bars of the
experimental points in the Figs.~\ref{figIII:dPhi-spectra}
and~\ref{figIII:dR-spectra}, as well as in the plots presented in
the experimental paper. Taking this large uncertainty into account,
it is reasonable to consider the overall normalization of the
cross-section to be a free parameter, which is also the case in MC
simulations presented in the Ref.~\cite{bCMS}. Following this route
we find, that to obtain a very good agreement of the central curve
of our predictions both with the shape and normalization of all
experimental spectra we have to multiply all our predictions on the
universal factor $\simeq 0.4$. Since the major part of the reported
normalization uncertainty is due to the uncertainty in the
efficiency of identification of $B$-mesons, our finding seems to
support the assumption, that the $B$-meson reconstruction efficiency
is largely independent from the kinematics of the leading jet, and
in particular, from the value of $p_{TL}^{\min}$. In the plots
below, we show theoretical predictions multiplied by the
above-mentioned factor, however our default result is also
compatible with experiment, if one takes into account full
experimental uncertainties and the scale-uncertainty of our
predictions.

In the Figs.~\ref{figIII:dPhi-spectra} and~\ref{figIII:dR-spectra}
we present the comparison of our predictions with $\Delta
\phi$ and $\Delta R$ spectra from the Ref.~\cite{bCMS}. Apart from
the above-mentioned overall normalization uncertainty, our model
does not contain any free parameters. To generate the gluon unPDF,
according to the Eq.~(\ref{eqI:KMR}) we use the LO PDFs from the
MSTW-2008 set~\cite{MSTW-2008}. We also use the value of
$\alpha_s(M_Z)=0.1394$ form the PDF fit. In both LO
(\ref{eqII:LO_subpr}) and NLO (\ref{eqII:NLO_subpr}) contrubutions
we set the renormalization and factorization scales to be equal to
the $p_T$ of the leading jet: $\mu_R=\mu_F=\xi p_{TL}$, where
$\xi=1$ for the central lines of our predictions, and we vary
$1/2<\xi<2$ to estimate the scale-uncertainty of our prediction,
which is shown in the following figures by the gray band. All
numerical calculations has been performed using the adaptive MC
integration routines from the \texttt{CUBA} library~\cite{CUBA},
mostly using the \texttt{SUAVE} algorithm, but with the cross-checks
against the results obtained by \texttt{VEGAS} and \texttt{DIVONNE}
routines.

The shape of measured distributions, both in $\Delta\phi$ and
$\Delta R$, agrees with our theoretical predictions within the
experimental uncertainty. Also, our model correctly describes the
dependence of the cross-section on the $p_{TL}^{\min}$ cut.

\begin{figure}[p!]
  \begin{center}
  \includegraphics[width=0.45\textwidth]{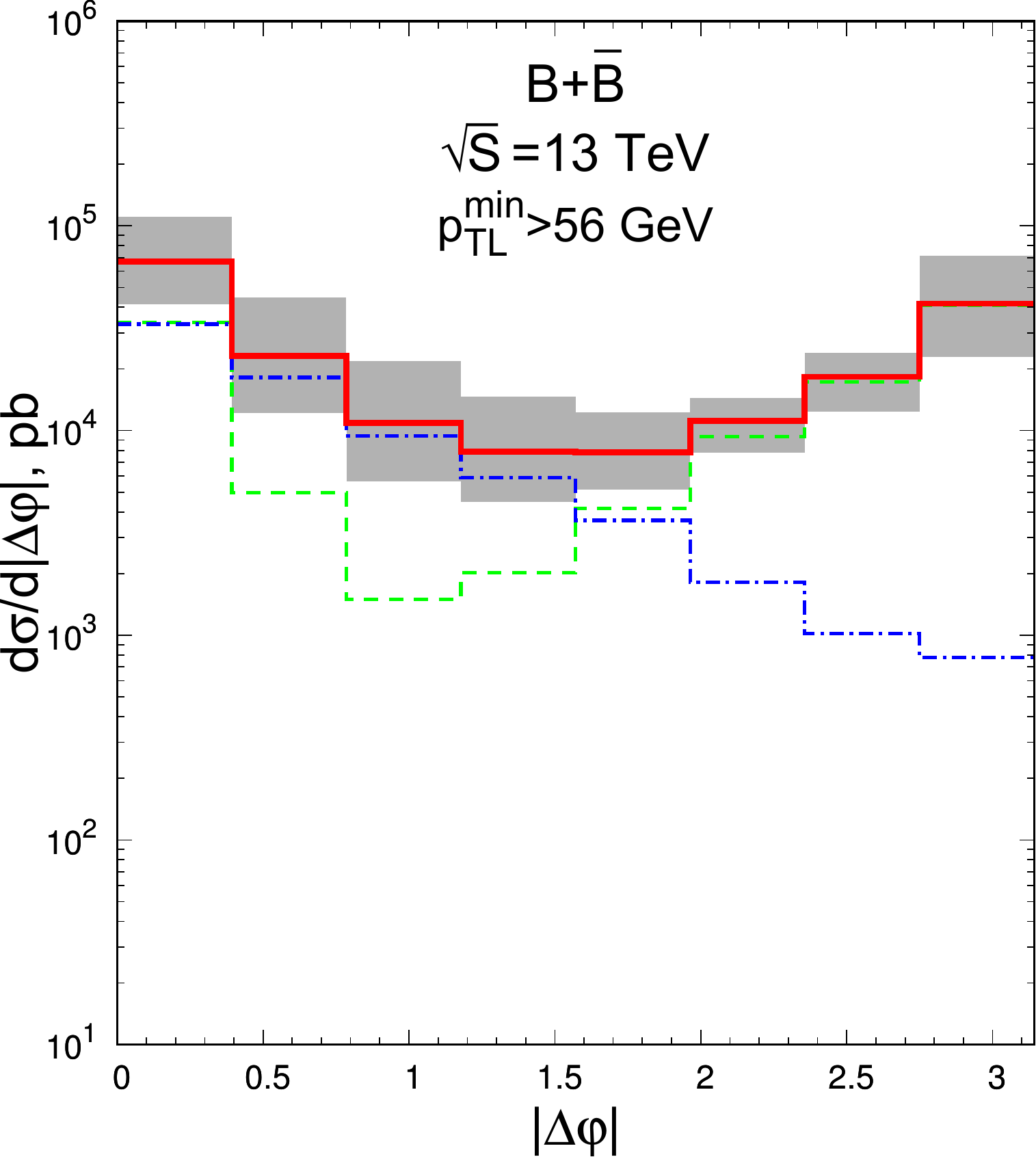}
  \includegraphics[width=0.45\textwidth]{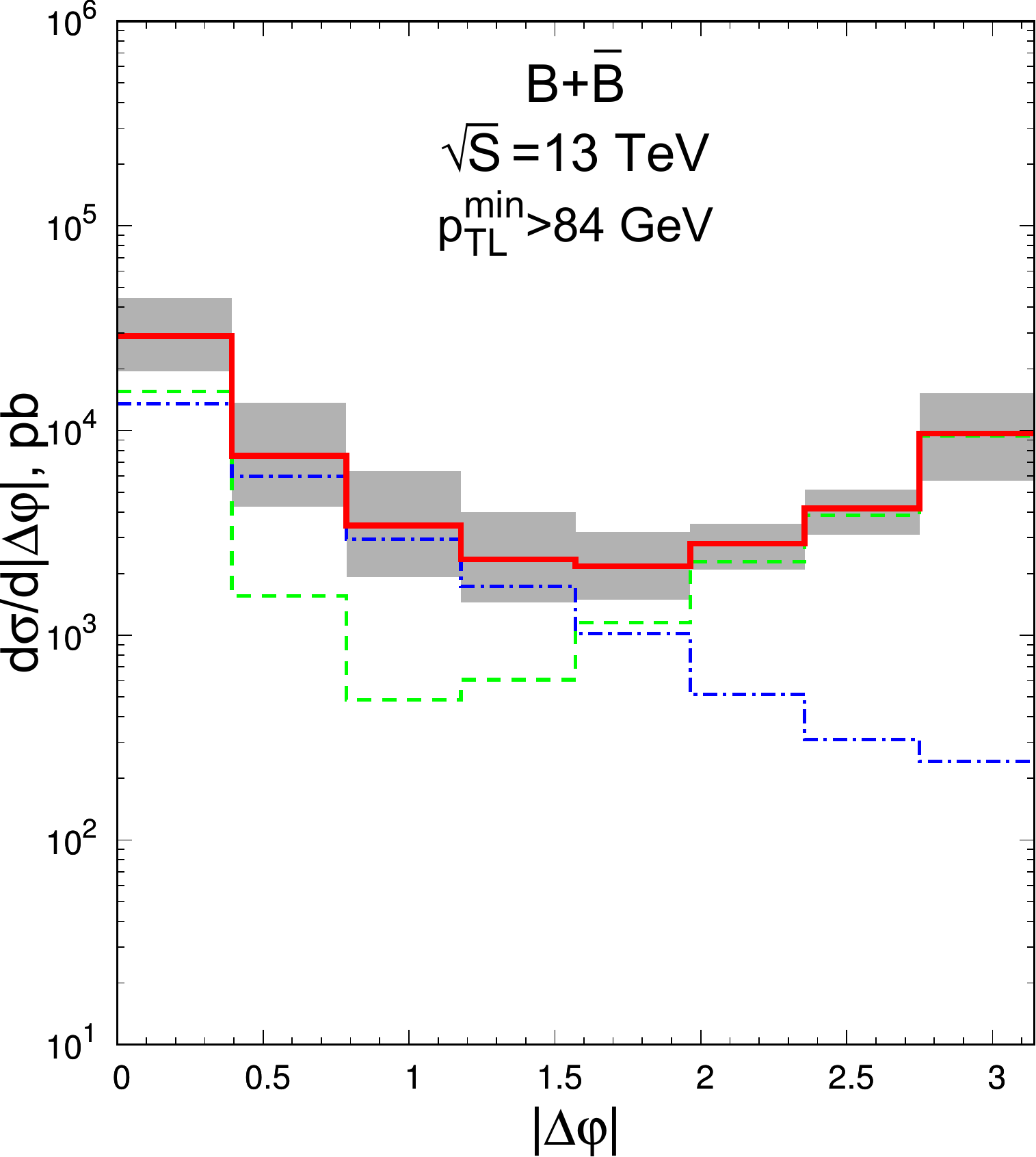} \\
  \includegraphics[width=0.45\textwidth]{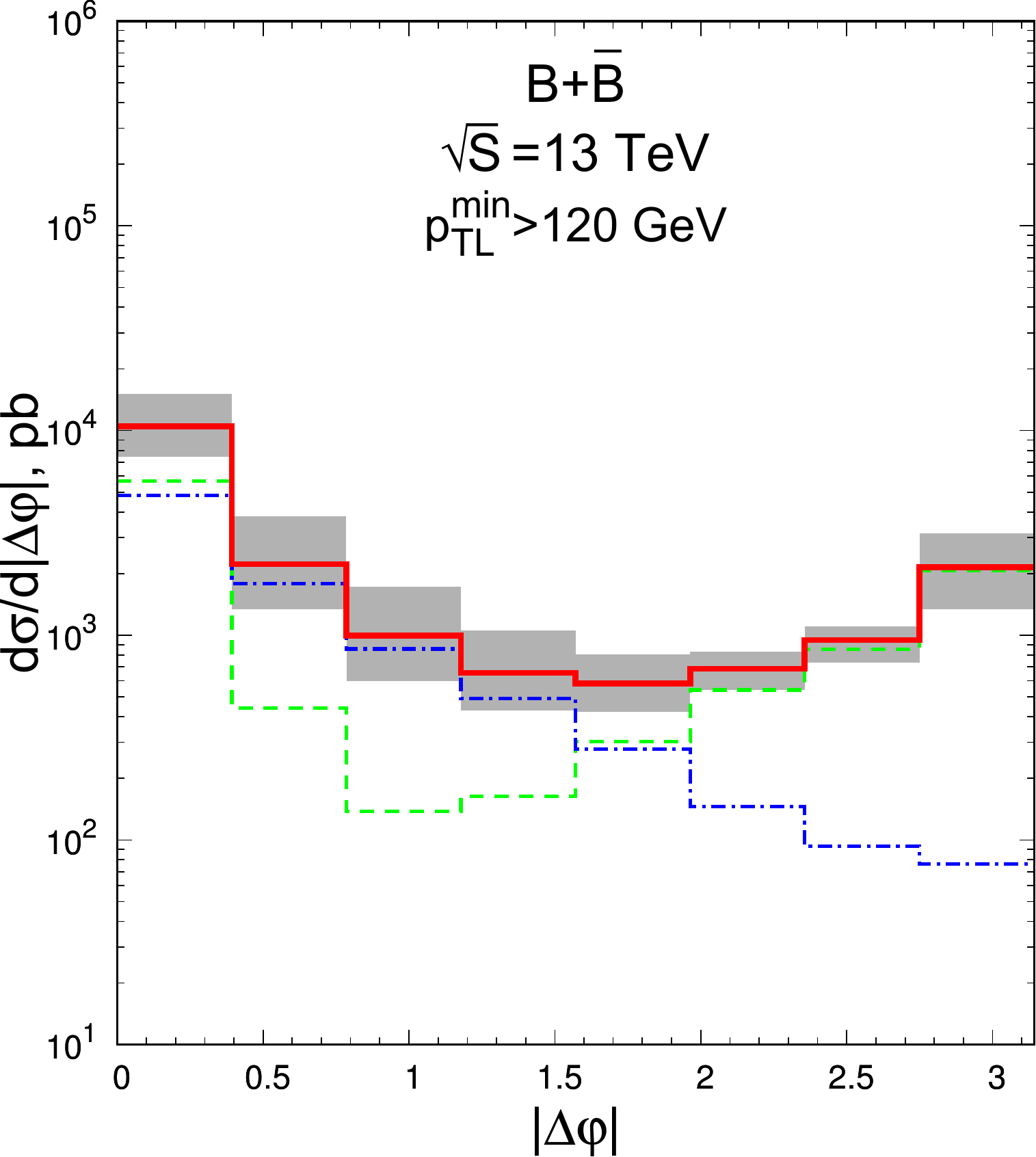}
  \end{center}
  \caption{Predictions for the $d\sigma/d\Delta\phi$-spectra $\sqrt{S}=13$ TeV for the same kinematic cuts as in the Ref.~\cite{bCMS}.
   Notation for the histograms is the same as in the Fig.~\ref{figIII:dPhi-spectra}.  \label{figIII:dPhi-spectra-13}}
  \end{figure}

\begin{figure}[p!]
  \begin{center}
  \includegraphics[width=0.45\textwidth]{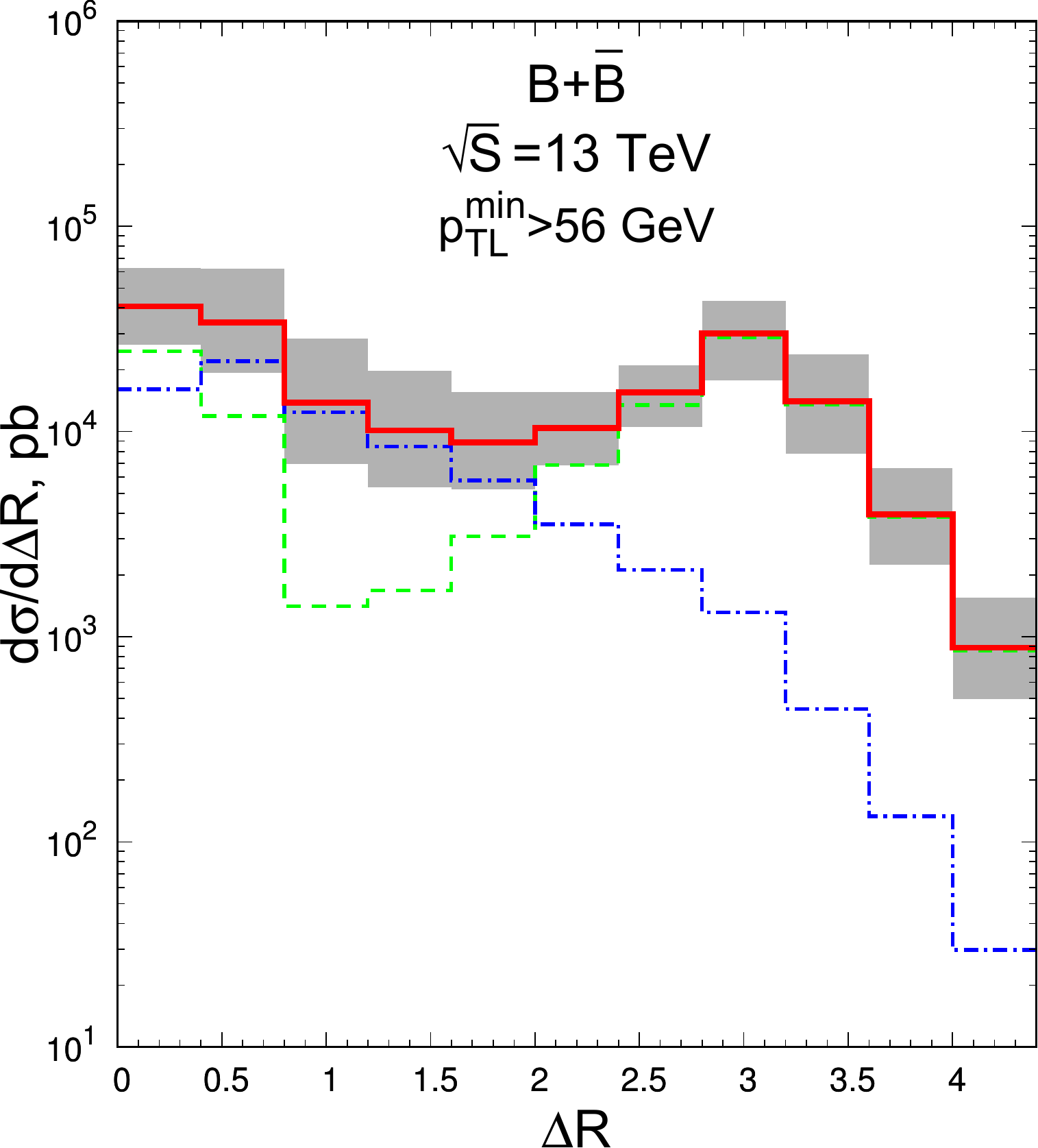}
  \includegraphics[width=0.45\textwidth]{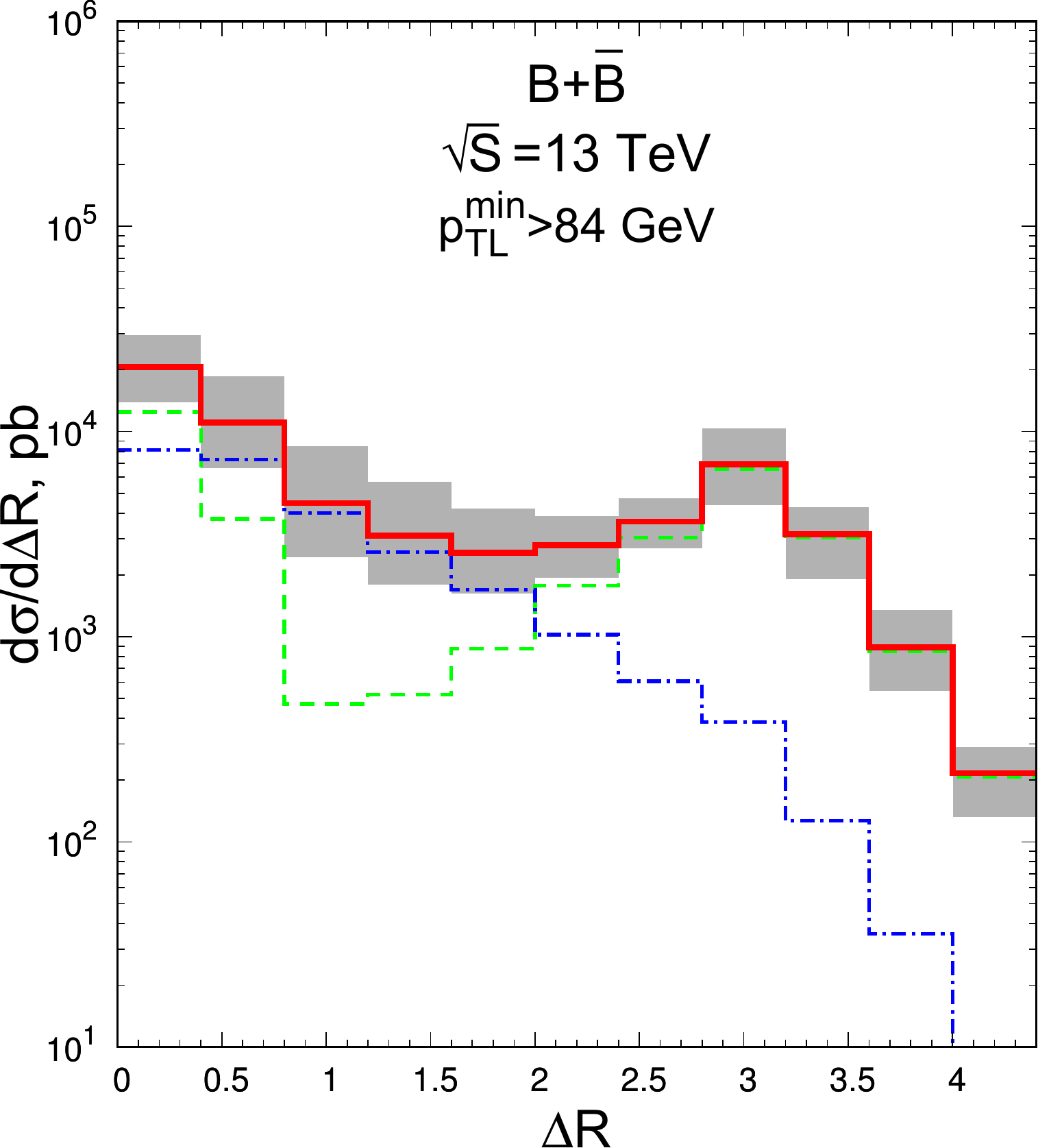} \\
  \includegraphics[width=0.45\textwidth]{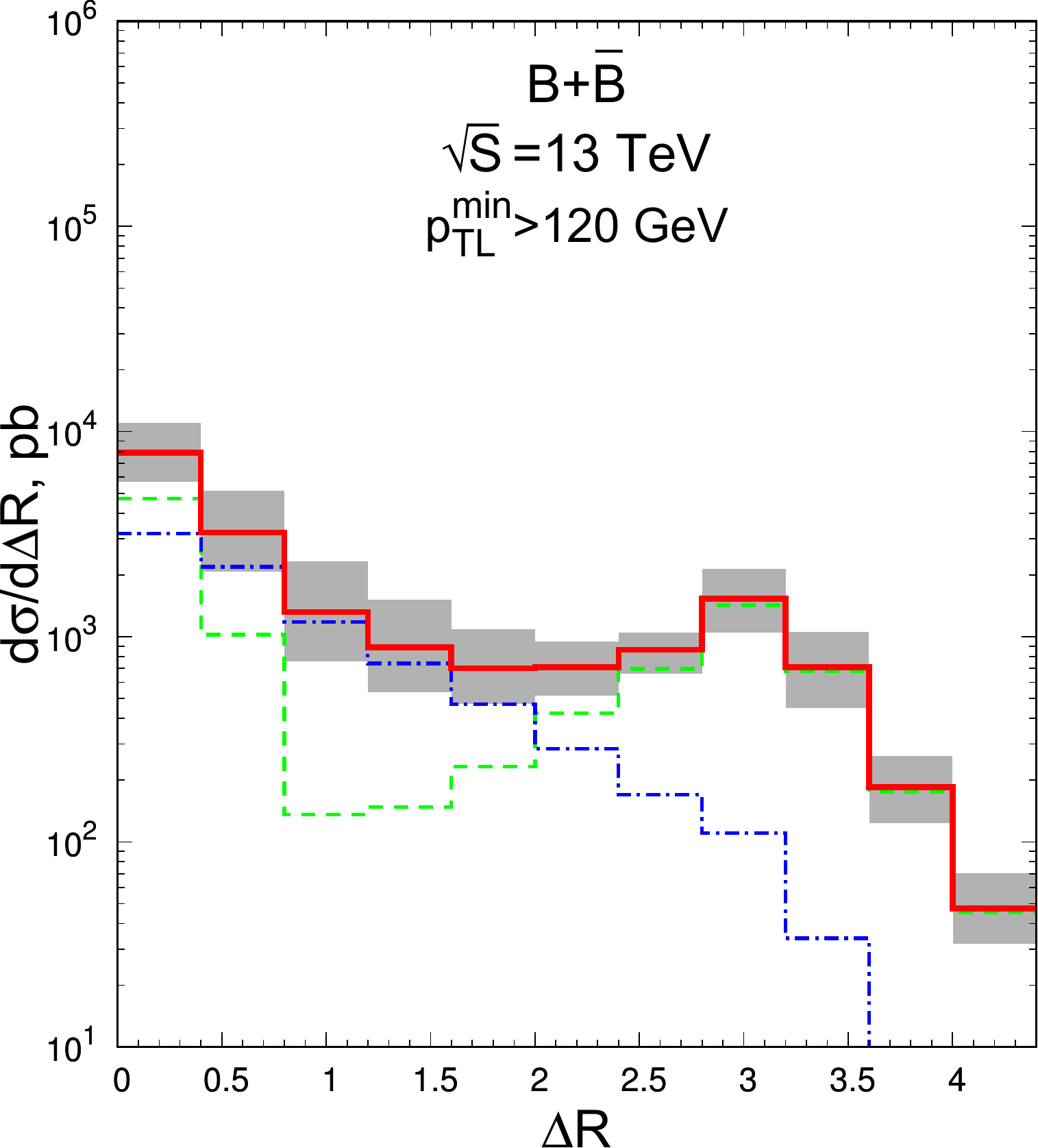}
  \end{center}
  \caption{Predictions for the $d\sigma/d\Delta R$-spectra $\sqrt{S}=13$ TeV for the same kinematic cuts as in the Ref.~\cite{bCMS}.
   Notation for the histograms is the same as in the Fig.~\ref{figIII:dPhi-spectra}.  \label{figIII:dR-spectra-13}}
  \end{figure}

\begin{figure}[p!]
  \begin{center}
  \includegraphics[width=0.45\textwidth]{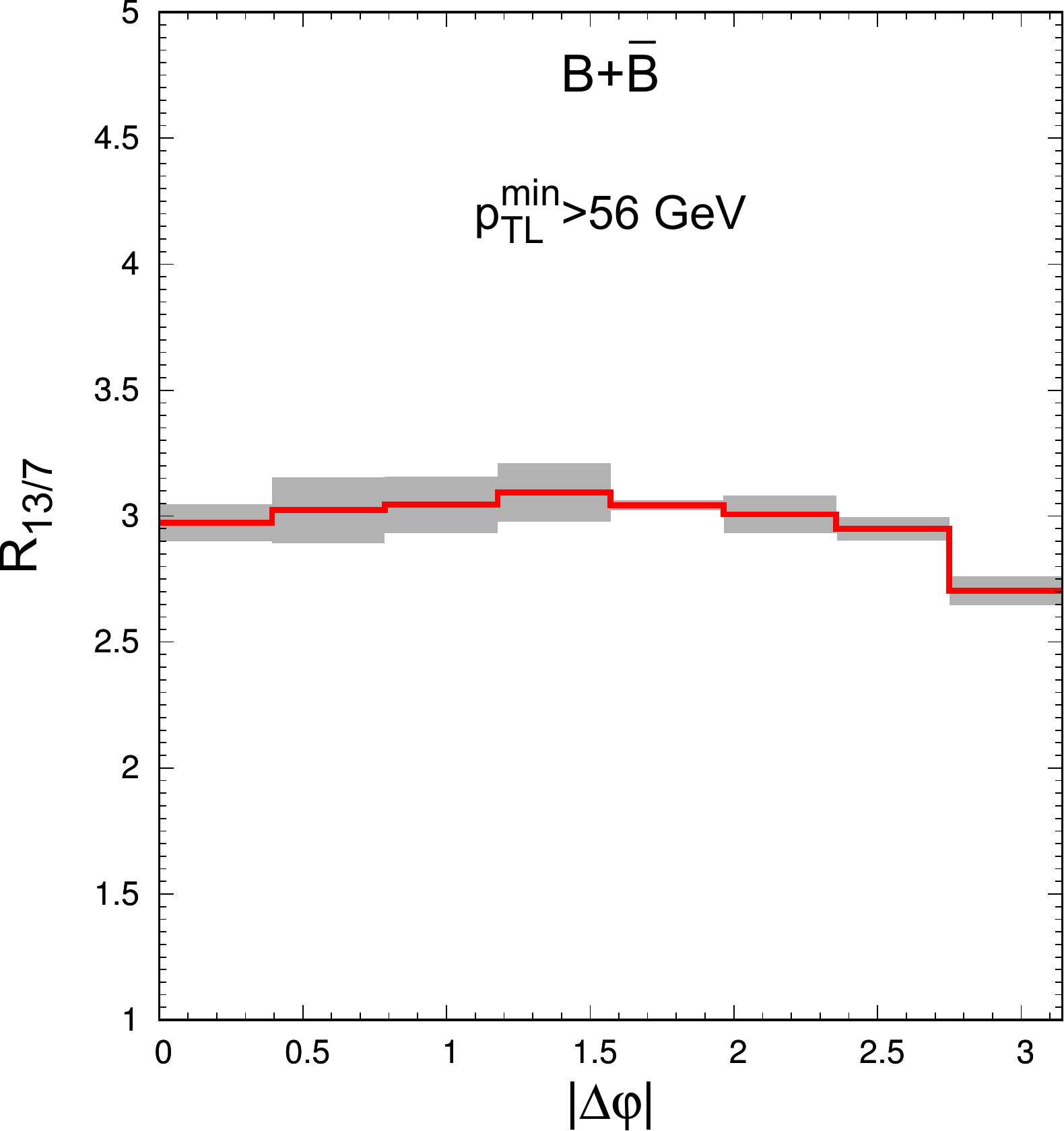}
  \includegraphics[width=0.45\textwidth]{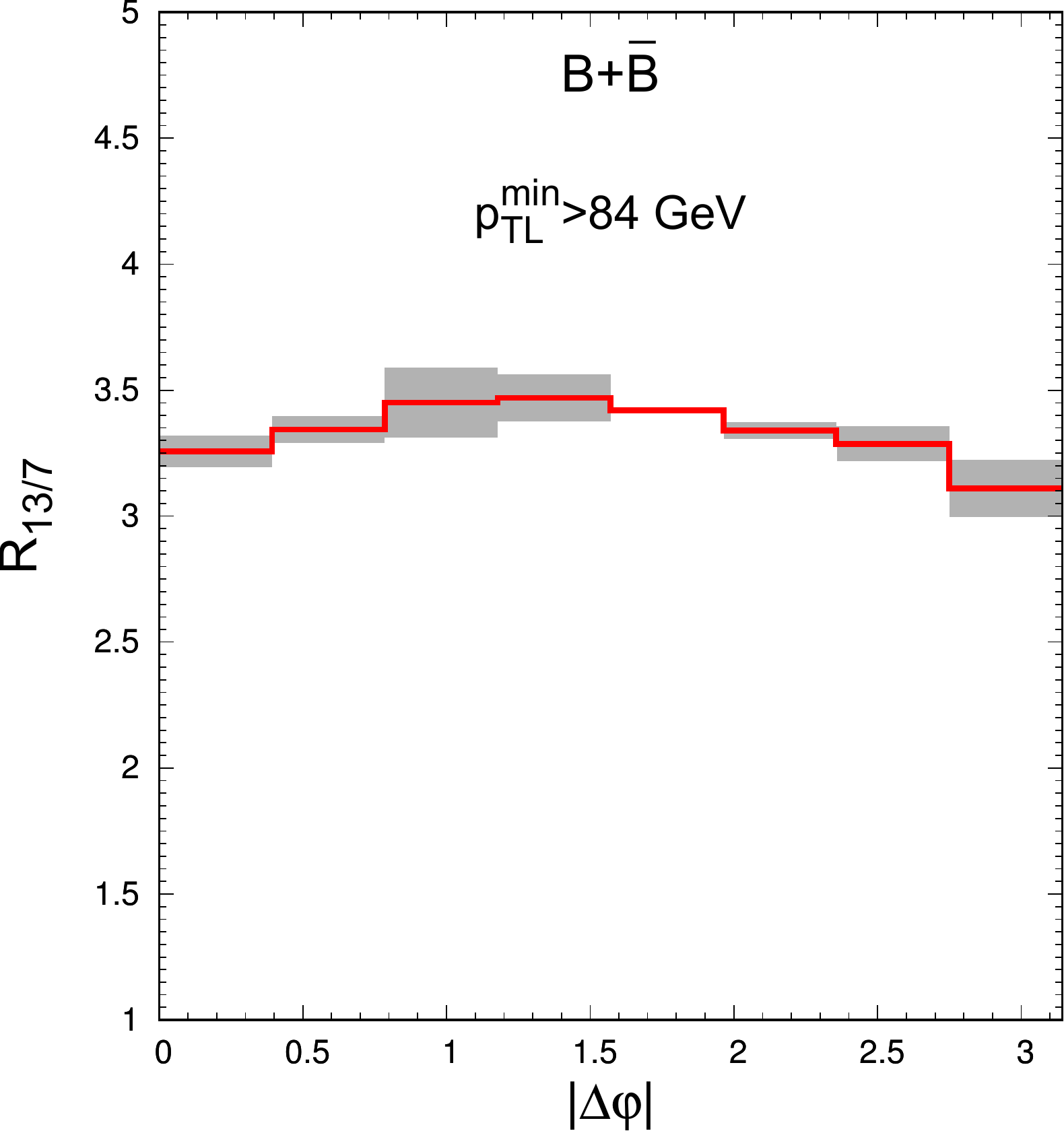} \\
  \includegraphics[width=0.45\textwidth]{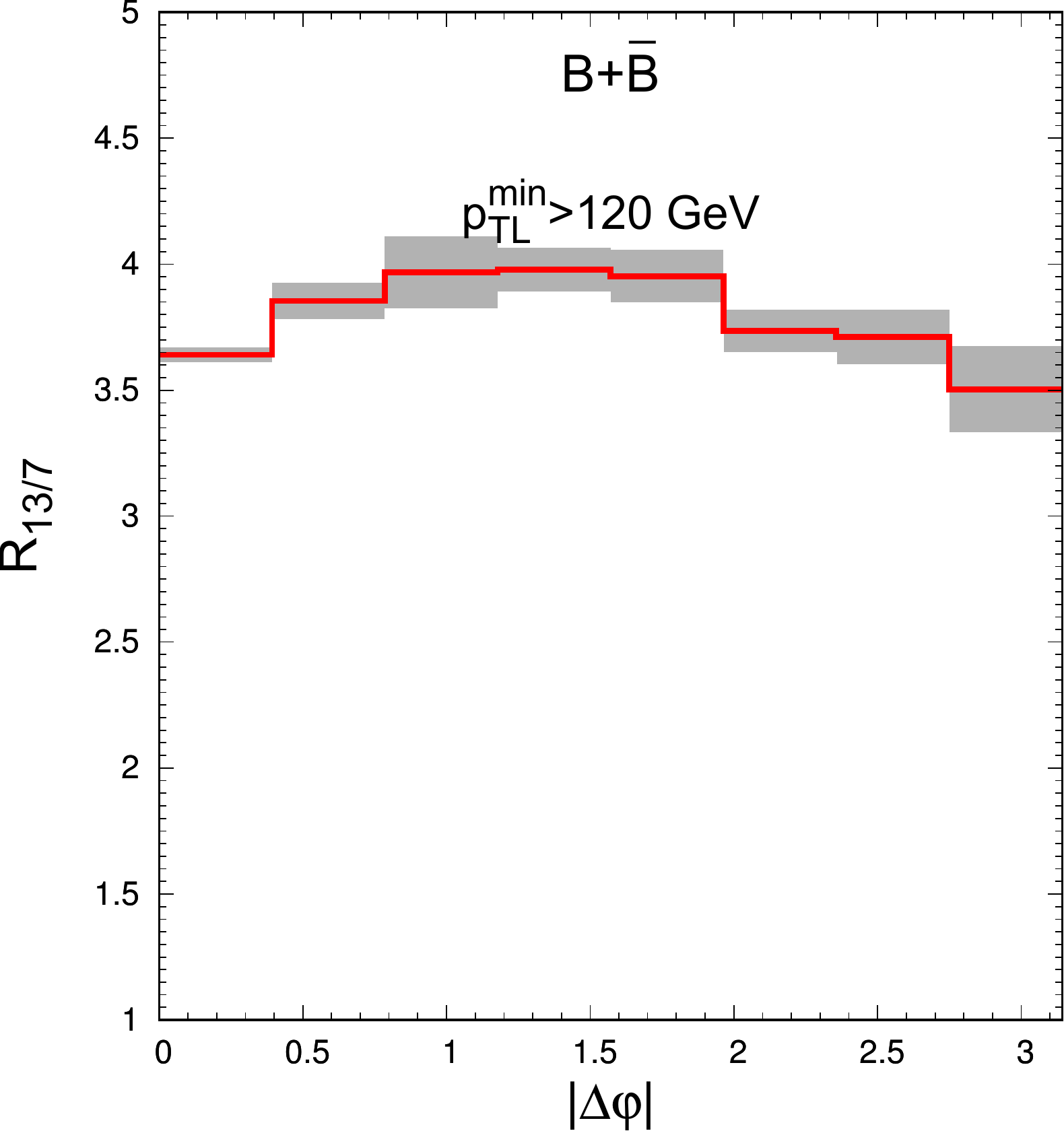}
  \end{center}
  \caption{Predictions for the ratio of $d\sigma/d\Delta\phi$-spectra at $\sqrt{S}=13$ TeV and $\sqrt{S}=7$ TeV
  for the same kinematic cuts as in the Ref.~\cite{bCMS}.  \label{figIII:dPhi-ratio}}
  \end{figure}

\begin{figure}[p!]
  \begin{center}
  \includegraphics[width=0.45\textwidth]{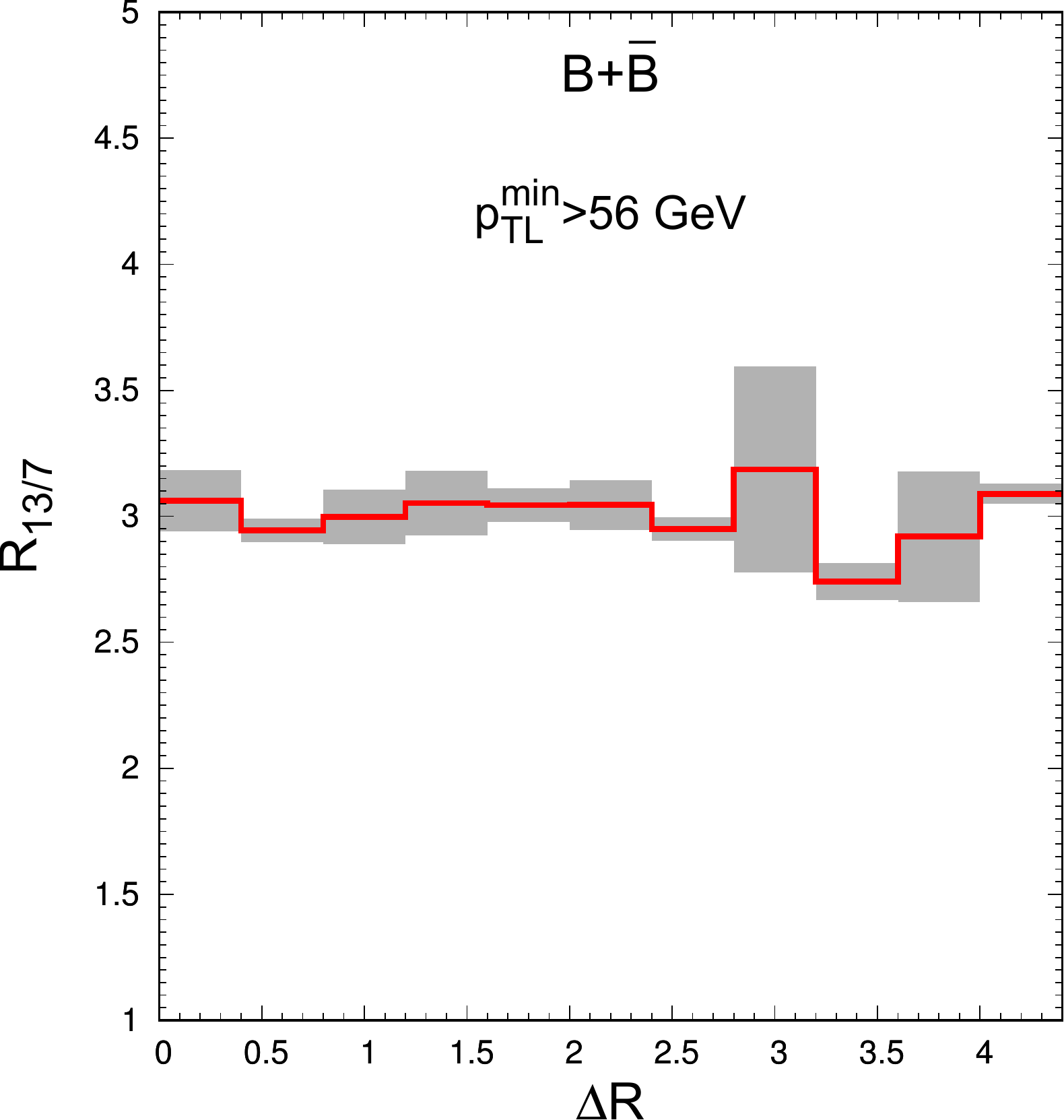}
  \includegraphics[width=0.45\textwidth]{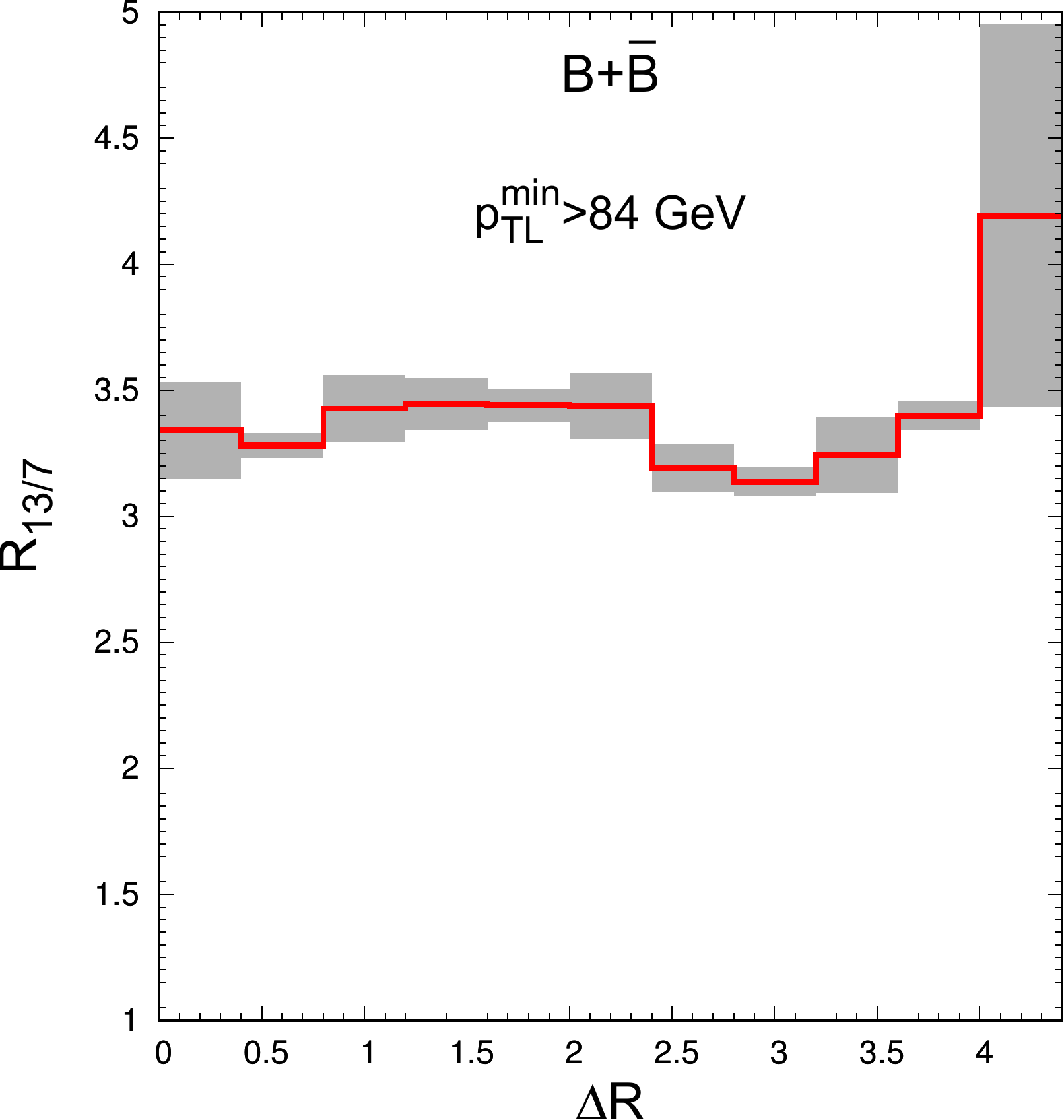} \\
  \includegraphics[width=0.45\textwidth]{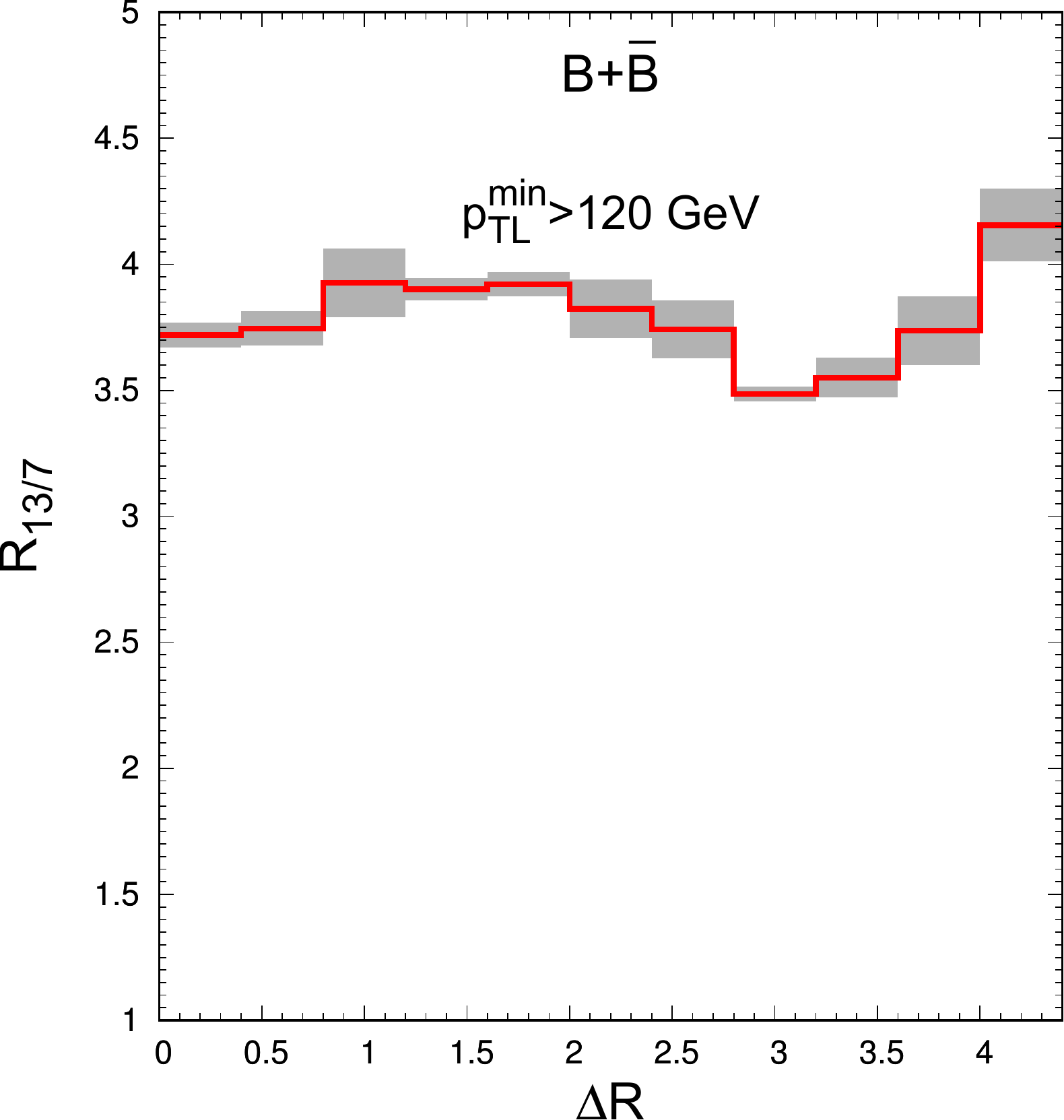}
  \end{center}
  \caption{Predictions for the ratio of $d\sigma/d\Delta R$-spectra at $\sqrt{S}=13$ TeV and $\sqrt{S}=7$ TeV for the same
   kinematic cuts as in the Ref.~\cite{bCMS}. \label{figIII:dR-ratio}}
  \end{figure}

Our predictions for the $d\sigma/d\Delta\phi$ and $d\sigma/d\Delta
R$ spectra at $\sqrt{S}=13$ TeV are presented in the Figs.
\ref{figIII:dPhi-spectra-13} and \ref{figIII:dR-spectra-13} for the
same kinematic cuts as in the Ref.~\cite{bCMS}. Also in the
Figs.~\ref{figIII:dPhi-ratio} and \ref{figIII:dR-ratio} we provide
predictions for the ratios of $\Delta\phi$ and $\Delta R$ spectra at
different energies, as it was proposed in the Ref.~\cite{ratios}.
The primary advantage of such observable is, that the theoretical
scale-uncertainty mostly cancels in the ratio, leading to the more
precise prediction. The residual $\Delta\phi$ and $\Delta R$
dependence of the ratio arises in the inerplay between the
$x$-dependence of PDFs and the dynamics of emissions of additional
hard radiation, therefore probing the physics of interest for PRA.
Measurements of such observables at the LHC will present an
important challenge for the state-of-the-art calculations in
perturbative QCD and tuning of the MC event generators.

\section{Conclusions}
\label{sec:conclusions}

In the present paper, the example of $B\bar{B}$-azimuthal
decorrelations is used to show, how the contributions of $2\to 2$
and $2\to 3$ processes in PRA can be consistently taken together to
describe multiscale correlational observables in a presence of
experimental constraints on additional QCD radiation. Our numerical
results agree well with experimental data of the Ref.~\cite{bCMS},
up to a common normalization factor. The predictions for
$\sqrt{S}=13$ TeV are provided. Also the foundations of the Parton
Reggeization Approach has been reviewed in the Sec.~\ref{sec:LO} and
the relation of PRA with collinear and Multi-Regge limits of
scattering amplitudes in QCD is higlighted.

\section{Acknowledgements}

We are grateful to A. van Hameren for help in comparison of squared
amplitudes obtained in the PRA with ones obtained using recursion
techniques of the Refs.~ \cite{hameren3, katie}. The work  was
supported by the Ministry of Education and Science of Russia under
Competitiveness Enhancement Program of Samara University for
2013-2020, project 3.5093.2017/8.9. The Feynman diagrams in the
Figs.~\ref{figI:FeynmanRules}, \ref{figI:FeynmanRules2} and
\ref{figI:MRK_ampl} were made using
\texttt{JAXODRAW}~\cite{JaxoDraw}.

\clearpage

\clearpage

\clearpage

\clearpage

\clearpage

\end{document}